\documentclass[11pt]{article}
\usepackage[body={15.8cm,23cm}]{geometry}
\usepackage{epsfig,epsf}
\usepackage{amsmath}
\usepackage{amsthm}
\usepackage{amsfonts}
\usepackage{amssymb}
\usepackage{psfrag}
\usepackage{cite}
\usepackage{slashed}

\renewcommand{\theequation}{\arabic{section}.\arabic{equation}}

\renewcommand{\thetable}{\arabic{table}}

\def \tr {\mathop{\rm tr}\nolimits}

\def \Re {\mathop{\rm Re}\nolimits}

\newcommand \widebar [1] {\overline{#1}}

\newcommand \vev [1] {\langle{#1}\rangle}

\newcommand{\as}{\ifmmode\alpha_{\rm s}\else{$\alpha_{\rm s}$}\fi}
\newcommand{\asbar}{\ifmmode\bar{\alpha}_{\rm s}\else{$\bar{\alpha}_{\rm s}$}\fi}

\newcommand \CR {\mathcal{R}}

\newcommand{\balpha}{{\boldsymbol{\alpha}}}
\newcommand{\bbeta}{{\boldsymbol{\beta}}}
\newcommand{\bsigma}{{\boldsymbol{\sigma}}}

\newcommand{\bgamma}{{\boldsymbol{\gamma}}}

\newcommand{\bw}{{\boldsymbol{w}}}

\newcommand{\bomega}{{\boldsymbol{\omega}}}
\newcommand{\brho}{{\boldsymbol{\rho}}}
\newcommand{\blambda}{{\boldsymbol{\lambda}}}
\newcommand{\bmu}{{\boldsymbol{\mu}}}
\newcommand{\be}{{\boldsymbol{e}}}

\font\cmss=cmss12 
\def\inbar{\,\vrule height1.5ex width.4pt depth0pt}
\def\IC{\relax\hbox{$\inbar\kern-.3em{\rm C}$}}
\def\IZ{\relax{\hbox{\cmss Z\kern-.4em Z}}}
\def\IR{{\hbox{{\rm I}\kern-.2em\hbox{\rm R}}}}

\def\IP{{\hbox{{\rm I}\kern-.2em\hbox{\rm P}}}}
\def\II{\hbox{{1}\kern-.25em\hbox{l}}}

\newbox\lett\newdimen\lheight\newdimen\lwidth
\def\ontop#1#2{\setbox\lett=\hbox{#2}\lheight\ht\lett
\multiply\lheight by 12 \divide\lheight by 10\relax%
\lwidth\wd\lett \multiply\lwidth by 8 \divide\lwidth by 10\relax #2\kern-\lwidth%
\raise\lheight\hbox{{$\scriptstyle #1$}}\kern.1ex}

\def\inbar{\,\vrule height1.5ex width.4pt depth0pt}

\newtheorem{theorem}{Theorem}
\newtheorem{lemma}{Lemma}

\numberwithin{equation}{section}

\begin{document}

\begin{titlepage}

\vspace*{20mm}
\noindent
  {\LARGE \bf Factorization of  $\mathcal{R}-$matrix and Baxter $\mathcal{Q}-$operators
for  generic $sl(N)$ spin chains.
 }

\vspace{5mm}
\noindent
{\it \large  Sergey \' E. Derkachov~$^\dag$} and
{\it \large  Alexander N. Manashov~$^{\ddag\S}$}
\\
\vspace*{0.1cm}

\noindent
$^\dag$~{\it St.Petersburg Department of Steklov
Mathematical Institute of Russian Academy of Sciences,
Fontanka 27, 191023 St.-Petersburg, Russia.}

E-mail:~derkach@euclid.pdmi.ras.ru
\vskip 5mm

\noindent
{$^\ddag$~\it
Institute for Theoretical Physics, University of  Regensburg,
D-93040 Regensburg, Germany.\\
$^\S$~
Department of Theoretical Physics,  Sankt-Petersburg  University,
St.-Petersburg, Russia.
}

 E-mail:~alexander.manashov@physik.uni-regensburg.de

\setcounter{footnote} 0

\noindent
\vskip 1cm

\hskip 10mm
\begin{minipage}{14cm}
{\small {\bf Abstract:} We develop an approach for constructing
the Baxter $\mathcal{Q}-$operators for  generic $sl(N)$ spin
chains. The key element of our approach is the possibility to
represent a solution of the the Yang Baxter equation in the
factorized form. We prove that such a representation holds for a
generic $sl(N)$ invariant $\mathcal{R}-$operator and find the
explicit expression for the factorizing operators. Taking trace of
monodromy matrices constructed of the factorizing operators one
defines a family of commuting (Baxter) operators on the quantum
space of the model. We show that a generic transfer matrix
factorizes into the product of $N$ Baxter $\mathcal{Q}-$operators
and  discuss an application of this representation for a
derivation of functional relations for transfer matrices. }
\end{minipage}
\vskip1cm

\end{titlepage}

{\tableofcontents}

\newpage

\section{Introduction}
The notion of an $\mathcal{R}-$matrix plays a key role in the theory
of lattice integrable systems. It is defined as a solution of the
Yang-Baxter equation (YBE)~\cite{Baxter}. The Quantum Inverse
Scattering Method (QISM)~\cite{FST,KuSk1,Skl,Fad} allows one to
relate an exactly solvable model with each solution of the YBE and
provides the methods of its analysis. These methods are the
Algebraic Bethe Ansatz (ABA)~\cite{FST,FaTa}, the method of Baxter
$Q-$operators~\cite{Baxter72} and Separation of Variables
(SoV)~\cite{Skl}. The most widespread and well studied of
them, ABA,  depends crucially on the existence of a pseudovacuum
state in the Hilbert space of a model. This requirement holds for
a majority of the models which found applications in  statistical
mechanics and quantum field theory. Systems like the Toda
chain~\cite{Gutzwiller:1981by} or noncompact spin
magnets~\cite{SL2C}, for which the ABA method is not applicable,
can be analyzed with the method of Baxter $Q-$operators and SoV.
Unfortunately, these methods are less developed in comparison with
 ABA and their application was so far restricted  to  models with low rank symmetry
groups. The main obstacle for the effective use of the method of
Baxter $Q-$operators is the absence of a regular method of their
construction.  Nevertheless, in recent time there was a certain
progress in the field. The Baxter operators are known now for a
number of models, see
Refs.~\cite{GP92,BLZ,Volkov,BLZ-II,AF96,BLZ-III,Hikami,SDQ,Pronko,SL2C,BHK,DKM-I,RW,KMS,DKM-II,Korff04,Korff06,Korff-c,Bytsko,BDKM-21,DM06,Kojima,BTsuboi}.
Special attention was paid to the analysis of the spin chain
models with $sl(n)$ or $U_q(\widehat{sl}(n))$ symmetry and their
supersymmetric extensions. In the approach developed by
V.Bazhanov, S.Lukyanov and A.Zamolodchikov~\cite{BLZ} the Baxter
$\mathcal{Q}-$operators are constructed as a trace of the special
monodromy matrix, the auxiliary space being an infinite
dimensional representation of the $q-$oscillator algebra. The method
was applied to the models associated with the affine quantum
algebras $U_q(\widehat{sl}(2))$~\cite{BLZ},
$U_q(\widehat{sl}(3))$~\cite{BHK} and
$U_q(\widehat{sl}(2|1))$~\cite{BTsuboi}. The construction relies
on the explicit solution for the universal
$\mathcal{R}-$matrix~\cite{KT1,KT2,KT3}. Since the latter is  rather
complicated, it causes technical problems if one  attempts to extend
the analysis to a general, $U_q(\widehat{sl}(N))$,
case~\cite{Kojima}.

The spin chain models associated with $sl(N=2,3)$ and $sl(2|1)$
algebras were analyzed in Refs.~\cite{SD-2,DM-I,DM-II,BDKM-21} by
a different method. It results in much the same functional
relations among the Baxter $\mathcal{Q}-$operators and  transfer
matrices as for the $q-$deformed models. This method relies heavily
upon  the special representation for the $\mathcal{R}-$operator.
Namely, it was shown in~\cite{SD} that the $sl(N=2,3)$ and
$sl(2|1)$ invariant $\mathcal{R}$ matrices on a tensor product of
generic representations can be represented as a product of two ($sl(2)$)
and three ($sl(3)$ and $sl(2|1)$) simpler operators. These operators possess a number of
remarkable properties which are easily translated into the
properties of the corresponding transfer matrices. There is no
doubt that the factorized representation for
$\mathcal{R}-$operator exists for a general $N$. 
The factorizing operators were constructed in the explicit form in Ref.~\cite{DM06}
for an invariant $\mathcal{R}-$operator acting on the tensor
product of the principal continuous  series representations of
$SL(N,\mathbb{C})$ group.
The aim of the
present paper is to show that the factorization holds for  an
invariant $\mathcal{R}-$operator on the tensor product of generic
(infinite dimensional) highest weight representations of the
$sl(N)$ algebra (Verma modules).

The Verma
module over $sl(N)$ algebra can be realized as a vector space of polynomials of $N(N-1)/2$ variables of
an arbitrary degree,~see sect.~\ref{roperator}.
We will use this realization throughout the paper.
The defining equations for the factorizing operators on Verma modules are too
complicated to be solved directly for a general $N$.
To find  a solution we will use the results of
Ref.~\cite{DM06}  where the  invariant $\mathcal{R}$ operator
acting on the tensor product of the
principal series representation of $SL(N,\mathbb{C})$
group was constructed.
This construction uses  the properties of the intertwining operators for the
principal series representations and naturally gives rise to the factorized form of the
$\mathcal{R}-$operator. The building blocks for the $\mathcal{R}-$operator are simple integral
operators  defined on  functions from $L^2(Z\times Z)$, where $Z$ is the group of 
lower unitriangular matrices.
We  try to interpret  these operators as  operators on  Verma modules.
Below we show that such interpretation is possible and find explicit expressions for the
factorising operators.

As an illustration let us briefly consider  the simplest case  of the $SL(2,\mathbb{C})$  group.
The $SL(2,\mathbb{C})$ invariant $\mathcal{R}-$operator
acts on the tensor product of two unitary principal series
representations, $T^{(s_1,\bar s_1)}\otimes  T^{(s_2,\bar s_2)}$,
$$
[T^{(s,\bar s)}(g) f](z)=(cz+d)^{-2s}(\bar c\bar z+\bar d)^{-2\bar s}
f\left(\dfrac{az+b}{cz+d}\right), $$
$\bar s=1-s^*$, $2(s-\bar s)=n$,
$f\in L^2(\mathbb{C})$.
It  can be represented in the  factorized form~\cite{DM06}
\begin{align}\label{ex1}
\mathcal{R}(u-v)= V(u_1-u_2)S(v_1-u_1)V(v_1-u_2)U(u_1-v_2)S(v_2-u_2)U(u_2-u_1)\,.
\end{align}
All operators $U,V,S$ depend on two spectral parameters, holomorphic and antiholomorphic ones,
the latter (antiholomorphic) is not  displayed explicitly,
i.e. $U(\lambda)\equiv U(\lambda,\bar\lambda)$, with $\lambda-\bar\lambda=n$ being integer.
In Eq.~(\ref{ex1}) we have put
 $u_1=u+1-s_1$, $u_2=u+s_1$, $v_1=v+1-s_2$, $v_2=v+s_2$,
and similarly for the barred parameters, $\bar
u_1=\bar u+1-\bar s_1$, etc.
The operators $U$ and $V$
are integral operators
\begin{align}\label{}
[U(\lambda)f](z,w)=&A(\lambda)\int d^2\xi \xi^{-1-\lambda}\,\bar
\xi^{-1-\bar\lambda}\,f(z-\xi,w)\,,
\notag\\
[V(\lambda)f](z,w)=&A(\lambda)\int d^2 \eta\, \eta^{-1-\lambda}\,\bar
\eta^{-1-\bar\lambda}\, f(z,w-\eta)\,,
\end{align}
(the factor $A(\lambda)$ is defined in Eq.~(\ref{A-factor}))
while the operator $S(\lambda)$ is  a multiplication operator
\begin{align}\label{}
[S(\lambda)f](z,w)=(z-w)^{-\lambda} (\bar z-\bar w)^{-\bar\lambda} f(z,w)\,.
\end{align}
For  $\lambda^*=-\bar\lambda$ all operators are unitary operators
with respect to the standard scalar product on
$L^{2}(\mathbb{C}\times\mathbb{C})$. It is clear that an
interpretation of  the operators $U$, $V$ and $S$ as  operators on
the product of Verma modules causes a lot of problems and hardly
possible at all. However, for  the products of  $U,V$ and $S$
operators, $R^{(1)}=VSV$ and $R^{(2)}=USU$, such interpretation
exists. Indeed,  the action of the operator $R^{(2)}$ on a test
function can be represented in the form~\footnote{The derivation
of Eq.~(\ref{r2}) makes use of the so-called star-triangle
relation~\cite{Isaev}, see e.g. Refs.~\cite{SL2C,DM06} for
details.}
\begin{align}\label{r2}
[R^{(2)} f](z,w)=A(u_2-v_2)\int d^2\xi\, [1-\xi]^{v_2-u_2-1}[\xi]^{u_1-v_2} f(\xi(z-w)+w,w)\,,
\end{align}
where we put for brevity $[\xi]^\lambda\equiv \xi^\lambda \bar \xi^{\bar\lambda}$. Let us
explore this integral for the case when $f(z,w)$ is a holomorphic polynomial in $z$ and $w$. It is
clear that the result is a polynomial again provided that the integrals
\begin{align}\label{Ik}
I_k(\alpha,\beta)=\int d^2\xi\, \xi^{k}\,[1-\xi]^{-\alpha}[\xi]^{-\beta}\,,
\end{align}
where $\alpha=1+u_2-v_2$ and $\beta=v_2-u_1$,
converge for an arbitrary $k$ in some region of the parameters $\alpha$ and $\beta$. Let
us assume that $\bar\alpha-\alpha=n>0$ and  $\bar\beta-\beta=m>0$.
The integral~(\ref{Ik}) converges in vicinity of the singular points $\xi=0,1,\infty$ if
$\Re{\alpha}<1$, $\Re{\beta}<1$ and $\Re{\bar\alpha+\bar\beta}>1$, respectively.
These conditions  can  always  be satisfied. Thus Eq.~(\ref{r2}) defines an operator on the
product of Verma modules. It could be checked that this operator coincides with
the factorizing operator obtained in~\cite{SD}.  Next, the integral~(\ref{Ik}) is
an analytic function of $\alpha,\beta$,
$$
I_k(\alpha,\beta)=\pi(-1)^{n+m+k}\dfrac{\Gamma(n-\alpha)}{\Gamma(\alpha)}
\dfrac{\Gamma(m-\beta)}{\Gamma(\beta-k)}\dfrac{\Gamma(n+m+\alpha+\beta-1)}{\Gamma(2+k-\alpha-\beta)}\,,
$$
hence one can extend the
domain of definition of operator~(\ref{r2}) to arbitrary $\alpha,\beta$.
Thus we have constructed the factorizing operator on the tensor product of $sl(2)$ Verma modules
starting from the solution for the $SL(2,\mathbb{C})$ case.

In what follows we  extend these arguments to a general case of $sl(N)$ invariant
$\mathcal{R}$ operator. Instead of a direct calculation of integrals (which becomes too
cumbersome) we accept another approach.
An operator on Verma module is determined by its matrix in some basis. However, such a 
description  is not very convenient. It is  preferable to describe
an operator by its kernel which is defined as follows, $A(z,w)=\sum_{kn} e_{k}(z) A_{kn} h_n(w)$,
where $e_k(z)$ and $h_n(w)$ are basis vectors in Verma module and its dual space.
The function $A(z,w)$ depends on the functional realization of the dual space, but as the
latter is fixed there is a one to one correspondence between an operator and its kernel.
Proceeding in this way we  derive the defining equation for the kernel of the factorizing
operator. Instead of solving this equation we show that the ``kernel'' of
$SL(N,\mathbb{C})$ operator calculated in some specific basis satisfies the same equation
and, hence, provides the kernel of the factorizing operator on Verma module.
The factorizing operators in the $SL(N,\mathbb{C})$ case possess a number of remarkable
properties.  The heuristic
arguments given above imply that  the operators obtained by
``$SL(N,\mathbb{C})\to sl(N)$ reduction'' should inherit
all these properties.

Taking  the trace of monodromy matrices constructed from the factorizing operators
one defines a family of commuting operators (Baxter $\mathcal{Q}-$operators) on the quantum
space of the model.
The trace is taken over an infinite dimensional space and, in a general, diverges for the model with
unbroken  $sl(N)$ symmetry.  The finiteness of the traces can be provided by introducing a regulating
factor~\cite{BLZ-III,BHK,Korff04,Korff06}  which breaks  $sl(N)$ symmetry down to its diagonal subgroup.
We prove that in this case the  traces for the Baxter operators converge absolutely.
Moreover, the Baxter operators can be identified with the generic $sl(N)$ transfer matrix with
a specially chosen auxiliary space.
Using the properties of the factorizing operators we show that  a generic transfer
matrix can be represented as a product of Baxter $\mathcal{Q}-$operators.
This
representation is quite helpful for the study of the functional relations among the
transfer matrices.
Though we will mainly discuss the homogeneous spin chains the approach is applicable to
the analysis of the inhomogeneous spin chain models.

The paper is organized as follows: In Section~\ref{ind} we recall some basic facts about
representations of $SL(N,\mathbb{C})$ group and fix  the notations. We  give also the  summary of
the results of  Ref.~\cite{DM06} which will be necessary for further discussion.
In Section~\ref{roperator} the $sl(N)$ invariant $\mathcal{R}-$operator
on the tensor product of two Verma modules is constructed. To handle some technical
problems we introduce an invariant bilinear form on a Verma module and describe an
operator by its kernel with respect to this form. Using this technique we find  the explicit
form of the factorizing operators in the $sl(N)$ case. We prove that these operators obey 
certain commutation relations. 
In Section~\ref{TmQ} we
construct the Baxter operators and study their properties. 
A generic  invariant transfer matrix is defined as the trace of the monodromy matrix over an
infinite-dimensional auxiliary space.
To ensure  convergence of the
trace we introduce a boundary operator 
which breaks the  $sl(N)$ symmetry of the transfer matrix to its diagonal subalgebra. 
We prove that the corresponding trace over an infinite-dimensional space exists and show that
a generic transfer matrix factorizes into a product of $N$ Baxter $\mathcal{Q}-$operators.
Section~\ref{summary} contains
concluding remarks.
The derivation of some technically involved results is presented in the Appendices.

%
          \section{$SL(N,\mathbb{C})$ invariant $\mathcal{R}$-operator:
           principal series representations
                    }
\label{ind}
%
%
To make further discussion self-contained we
give here a brief review of the principal series representations of the $SL(N,\mathbb{C})$
group, and then formulate the results of the Ref.~\cite{DM06} which will be used in
subsequent analysis.

           \subsection{Principal series representation of $SL(N,\mathbb{C})$}

The unitary principal series representations of the group $SL(N,\mathbb{C})$ can be
realized on the space of functions on the group of lower triangular $N\times N$
matrices~~\cite{Gelfand,SG}. Namely,
let $Z_-(Z_+)$ and $H_+(H_-)$  be  the groups of  lower (upper) unitriangular
matrices and   upper (lower) triangular
matrices with  unit determinant, respectively,
\begin{align*}
{z}=
\begin{pmatrix}1&0&0&\ldots&0\\
z_{21}&1&0&\ldots&0\\
z_{31}&z_{32}&1&\ldots&0\\
\vdots&\vdots&\vdots&\ddots&\vdots\\
z_{N1}&z_{N2}&\ldots&z_{N,N-1}&1
\end{pmatrix}\in Z_-\,,&&
h=
\begin{pmatrix}
  h_{11} & h_{12} & h_{13}&\ldots & h_{1N} \\
  0 & h_{22} & h_{23}&\ldots & h_{2N} \\
  0 & 0 & h_{33}&\ldots & h_{3N} \\
  \vdots & \vdots & \vdots&\ddots &\vdots \\
  0 & 0 &0 &\ldots & h_{N,N}
\end{pmatrix}\in H_+\,.
\end{align*}
Almost any  matrix ${g}\in G=SL(N,\mathbb{C})$ admits the Gauss decomposition
${g}={z}\,{h}$. 
The element $z_1\in Z_-$ satisfying the condition
$g^{-1}\cdot z=z_1\,\cdot h\,$
will be denoted by $z\bar g$, so that
$g^{-1}z=z\bar g\cdot h$.
The homomorphism $g\to T^{{\balpha}}(g)$,
\begin{align}\label{Tg}
[T^{\balpha}(g)\,\Phi](z)=\balpha(h^{-1})\,\Phi(z\bar g)\,,
\end{align}
defines a principal series representation of the group $SL(N,\mathbb{C})$ on a suitable
space of functions on the group $Z_-$,
$\Phi(z)=\Phi(z_{21},\bar z_{21},z_{31},\bar z_{31},\ldots,z_{NN-1},\bar z_{NN-1})$ ~\cite{Gelfand,SG}.
The function $\balpha$ in Eq.~(\ref{Tg}) is the character of the group $H_+$,
\begin{align}\label{dh}
{\balpha}(h)=\prod_{k=1}^N
h_{kk}^{-\sigma_k-k}\,\bar
h_{kk}^{-\bar\sigma_k-k}\,.
\end{align}
Here  $\bar h_{kk}\equiv (h_{kk})^*$ is the complex conjugate of $h_{kk}$,
whereas in general  $\sigma_k^*\neq \bar \sigma_k$. We put
$\bsigma=(\sigma_1,\ldots,\sigma_N)$ and will sometimes use notation $T^\bsigma$ instead
of $T^\balpha$.
Since $\det h=1$ the function $\balpha(h)$ depends only on the differences
$\sigma_{k,k+1}\equiv\sigma_k-\sigma_{k+1}$   and can be rewritten in the form
\begin{align}\label{aDk}
\balpha(h)=&\prod_{k=1}^{N-1}
(\Delta_{k}(h))^{1-\sigma_{k,k+1}}
(\bar \Delta_{k}(h))^{1-\bar\sigma_{k,k+1}}
=\prod_{k=1}^{N-1}(\Delta_{k}(h))^{n_k} \, \vert \Delta_{k}(h)\vert^{2(1-\bar\sigma_{k,k+1})}\,,
\end{align}
where $n_k= \bar \sigma_{k,k+1}-\sigma_{k,k+1}$, $k=0,\ldots,N-1$
are  integer numbers, $n_k\in \mathbb{Z}$~\footnote{From now in, since each
variable $a$ comes along with
its  antiholomorphic twin $\bar a$ we will write down only holomorphic
variant of equations.}.
The function $\Delta_k(M)$ is defined by
\begin{align}\label{DeltaM}
\Delta_{k}(M)=\det M_k,
\end{align}
where the $k\times k$ matrix $ M_k$ is  $k$-th main minor of the matrix $ M$, and
$\bar\Delta_k(M)=(\Delta_k(M))^*$. That is
$\Delta_k(h)=\prod_{i=1}^k h_{ii}$  for $h\in H_+$.
We will assume that the parameters
$\sigma_k$ satisfy the restriction
\begin{align}\label{rs}
\sigma_1+\sigma_2+\ldots+\sigma_N=N(N-1)/2.
\end{align}
The operator $T^{\balpha}(g)$ is
a unitary operator on the Hilbert space $L^2(Z_-)$,
\begin{align*}
\vev{\Phi_1|\Phi_2}=\int Dz\,  (\Phi_1(z))^*\, \Phi_2(z)\,,&& Dz=\prod_{1\leq i<k\leq N} d^2 z_{ki}\,,
\end{align*}
if the character   $\balpha'(h)=\balpha(h)\prod_{k=1}^N |h_{kk}|^{2k}$ is a unitary
one, i.e. $\vert\balpha'\vert=1$.
This condition  holds if
$\sigma_{k,k+1}^*+\bar\sigma_{k,k+1}=0$ for $k=1,\ldots,N-1$, i.e.
\begin{align}\label{unitarity}
\sigma_{k,k+1}=-\frac{n_k}2+i\lambda_k\,, &&
\bar\sigma_{k,k+1}=\frac{n_k}2+i\lambda_k\,,&&k=1,2,\ldots,N-1\,,
\end{align}
where $n_k$ is integer and $\lambda_k$ is real.
The unitary principal series representation $T^{\balpha}$ is irreducible. Two representations
$T^{\balpha}$ and  $T^{\balpha'}$ are unitary equivalent if and only if the
corresponding parameters $(\sigma_1,\ldots,\sigma_N)$ and $(\sigma'_1,\ldots,\sigma'_N)$
are related  by a permutation, see for details Refs.~\cite{Gelfand,SG}.

\vskip 3mm
We will also need the explicit expression for the generators of infinitesimal $SL(N,\mathbb{C})$
transformations, which are defined in the standard way
\begin{align*}%
\big[T^\balpha(\II+\sum\nolimits_{ik}\epsilon^{ik} \mathcal{E}_{ki})\,\Phi\Big](z)=
\Phi(z)+\sum\nolimits_{ik}\left(\epsilon^{ik}\,E_{ki}+\bar\epsilon^{ik}\,\bar E_{ki}\right)\,
\Phi(z)+O(\epsilon^2)\,.
\end{align*}
Here $\mathcal{E}_{ik}$, ($1\leq i,k\leq N$), are the generators in the
fundamental representation of the  $SL(N,\mathbb{C})$ group,
\begin{align}\label{fund}
(\mathcal{E}_{ik})_{nm}=\delta_{in}\delta_{km}-\frac1N\delta_{ik}\delta_{nm}.
\end{align}
The generators
$E_{ik}(\bar E_{ik})$ are  linear differential operators in the variables $z_{mn}(\bar
z_{mn})$, ($1\leq n<m\leq
N$) which satisfy the  commutation relation
\begin{align}\label{comm}
[E_{ik} \,,E_{nm}]=\delta_{kn}\, E_{im}-\delta_{im}\, E_{nk}\,.
\end{align}
The generators $E_{ik}$ admit the following representation~\cite{DM06}
\begin{align}\label{DE}
E_{ik}=-\sum_{m\leq n}z_{km}\,\Bigl(D_{nm}+\delta_{nm}\,\sigma_m\Bigr)\,(z^{-1})_{ni}\,.
\end{align}
Here $D_{nm}$, $n>m$ are the generators of the right shifts, $\Phi(z)\to \Phi(z z_0)$,
\begin{align}\label{rshifts}
\Phi\Big(z\left(\II+\sum\nolimits_{k>i}\epsilon^{ik}\mathcal{E}_{ki}\right)\Big)=
\Big(1+\sum\nolimits_{k>i}\left(\epsilon^{ik}D_{ki}+\bar\epsilon^{ik}\bar D_{ki} \right)
+O(\epsilon^2)\Big)\,\Phi(z)\,.
\end{align}
Equation~(\ref{DE}) can be written in the matrix form
\begin{align}\label{EzD}
E=-z\left(\sigma+D\right)z^{-1}\,,
\end{align}
where
\begin{align}\label{EDs}
E=\sum_{ik} E_{ik} e_{ki},,&& D=\sum_{i>k} D_{ik} e_{ki}\,,&&\sigma=\sum_k \sigma_k e_{kk}
\end{align}
 and  the matrices $e_{nm}$ ($n,m=1,\ldots,N$) form the
standard basis in the space $\text{Mat}(N\times N)$,
$(e_{nm})_{ik}=\delta_{in}\delta_{mk}\,.$
The generators $D_{ki}$  satisfy  the same commutation relation as
$E_{ki}$,~(Eq.~\eqref{comm})
$
[D_{ki} \,,D_{nm}]=\delta_{in}\, D_{km}-\delta_{km}\, D_{ni}\,
$
and commute with the generators of left shifts, $E_{ki}$, $k>i$.
An explicit expression for the
generators of left and right shifts reads
\begin{align}\label{DD}
E_{ki}=-\sum_{m=1}^{i} z_{im}\,\frac{\partial\phantom{z_{m}}}{\partial z_{km}}
=\sum_{m=k}^N \tilde z_{mk}\,\frac{\partial\phantom{z_{m}}}{\partial \tilde z_{mi}}
\,, \\
D_{ki}=-\sum_{m=1}^{i} \tilde z_{im}\frac{\partial\phantom{z_{m}}}{\partial{\tilde z_{km}}}
=\sum_{m=k}^N z_{mk}\,\frac{\partial\phantom{z_{m}}}{\partial z_{mi}}
\,,
\end{align}
where $\tilde z_{ki}=(z^{-1})_{ki}$ and  we recall that $z_{ii}=1$.
Let us notice here that the operator $D_{ki}$ depends on the variables in the $k-$th and
$i-$th columns of the matrix $z$, or  on the variables in the  $k-$th and
$i-$th rows of the inverse matrix $z^{-1}$.

\subsection{Coherent states}
In this subsection we describe  the  system of functions with  ``good'' transformation
properties with respect to $SL(N,\mathbb{C})$ transformations. This system will play a key role
for establishing relationship between $\mathcal{R}-$operators defined on different
spaces of functions.
Namely, we define
\begin{align}\label{}
\Delta^{\bsigma}(z,\alpha)=\prod_{k=1}^{N-1}(\Delta_k(\alpha z))^{-1+\sigma_{k,k+1}}
\,(\bar\Delta_k(\alpha z))^{-1+\bar \sigma_{k,k+1}}\,,
\end{align}
where $\Delta_k(M)=\det M_k$,~see Eq.~(\ref{DeltaM}). The function
$\Delta^\bsigma(z,g^{-1})$ is nothing else than the prefactor $\balpha$ in
Eq.~(\ref{Tg}), i.e. $\Delta^\bsigma$ is a
transformation of unity,
\begin{align}
\Delta^\bsigma(z, g^{-1})=T^\balpha(g)\cdot1=\balpha(h^{-1}(z,g))\,.
\end{align}
We will refer to  the function $\Delta^\bsigma(z,\alpha)$, with $\alpha$ being an upper
triangular matrix, \mbox{$\alpha\in Z_+$},
as a coherent state.
For an unitary representation 
the system of coherent states \mbox{$\{\Delta^\bsigma(z,\alpha), \alpha\in Z_+\}$}
is a complete orthogonal system in $L^2(Z)$.
Indeed,  it is easy to verify that the integral operator~$\mathfrak{D}$ defined as
\begin{align}\label{}
[\mathfrak{D}\varphi](z)=\int D\alpha \,\Delta^\bsigma(z,\alpha)\,\varphi(\alpha)\,,
&& \alpha\in Z_+,\qquad D\alpha=\prod_{1\leq i<k\leq N} d^2 \alpha_{ik}
\end{align}
intertwines the unitary representations
$T^{\balpha}$ and $\widetilde T^{\bgamma}$,
\begin{align}\label{TDT}
T^{\balpha}(g)\,\mathfrak{D}=\mathfrak{D}\,\widetilde T^{\bgamma}(g)\,.
\end{align}
The representation $\widetilde T^{\bgamma}$ is defined on  functions on the
group $Z_+$,
\begin{align}\label{Ttilde}
[\widetilde T^{\bgamma}(g)f](\alpha)=\bgamma(h)\, f(\alpha\bar g)\,, &&
\bgamma(h)=&\prod_{k=1}^{N-1}
\left(\Delta_{k}(h)\right)^{-1-\sigma_{k,k+1}}
\left(\bar \Delta_{k}(h)\right)^{-1-\bar\sigma_{k,k+1}}\,.
\end{align}
Here  for $\alpha\in Z_+$ and $g\in SL(N,\mathbb{C})$ 
we  put $\alpha\cdot g=h\cdot \alpha\bar g$, $h\in H_-$.
It follows from Eq.~(\ref{TDT}) that the operator $\mathfrak{D}\mathfrak{D}^\dagger$ commutes with all
operators $T^\balpha(g)$. Since $T^\balpha(g)$ is an operator-irreducible
representation~\cite{Gelfand} the operator
$\mathfrak{D}\mathfrak{D}^\dagger$ is proportional to the unit operator. Hence
\begin{align}\label{}
\int Dz\,
\Delta^\bsigma(z,\alpha)\,\overline{\Delta^\bsigma(z,\alpha')}=c_N(\bsigma)
\prod_{1\leq k<i\leq
N}\delta^2(\alpha_{ik}-\alpha'_{ik})\,,
\end{align}
where $\delta^2(z)=\delta(x)\delta(y)$  for complex $z=x+iy$.
For the normalization factor we obtained
\begin{align*}\label{}
c_N(\bsigma)=\prod_{1\leq k<i\leq N}\frac{\pi^2}{|\sigma_i-\sigma_k|^2}\,.
\end{align*}
As was mentioned above  coherent states have  good transformation
properties. Namely, one easily derives
\begin{align}\label{ala}
\balpha^{-1}(h_+(z,g))\,\Delta^\bsigma(z\bar
g,\alpha)=\balpha^{-1}(h_-(\alpha,g^{-1}))\,\Delta^\bsigma(z,\alpha\bar g^{-1})\,,
\end{align}
where, we recall,   $ g^{-1}\cdot z =z\bar g\cdot h_+(z,g)$, $\alpha\cdot
g^{-1}=h_-(\alpha,g^{-1})\cdot\alpha\bar g^{-1}$.
Equation~(\ref{ala}) can be brought into the form
\begin{align}\label{TtT}
T_z^{\bsigma}(g)\,\Delta^\bsigma(z,\alpha)=
\widetilde T_\alpha^{-\bsigma}(g^{-1})\,\Delta^\bsigma(z,\alpha)\,,
\end{align}
where the transformation $\widetilde T^{-\bsigma}(g)$ is given by Eqs.~(\ref{Ttilde}) with
the substitution $\sigma_k\to-\sigma_k$. Thus the coherent state
$\Delta^\bsigma(z,\alpha)$ satisfies the equation
\begin{align}\label{EwE}
\Big(E_{ik}^{(z)}+\widetilde E_{ik}^{(\alpha)}\Big)\,\Delta^\bsigma(z,\alpha)=0\,,
\end{align}
where $E_{ik}$ and $\widetilde E_{ik}$, $i,k=1,\ldots,N$,
 are the generators of the $sl(N)$ algebra in the
representations $T^\bsigma$ and $\widetilde T^{-\bsigma}$, respectively.

%
\subsection{Factorized form of $\mathcal{R}$ operator}
%
The Yang-Baxter  equation (YBE) is an operator equation which is a corner
stone of the theory of  integrable systems. It has the form
\begin{align}   \label{YB}
\CR_{12}(u-v)\CR_{13}(u-w)\CR_{23}(v-w)=\CR_{23}(v-w)\CR_{13}(u-w)\CR_{12}(u-v)\,.
\end{align}
The operators act on the tensor product  $\mathbb{V}_1\otimes\mathbb{V}_2\otimes\mathbb{V}_3$.
As usual, an operator with the indices $ik$ acts nontrivially on the space
$\mathbb{V}_i\otimes \mathbb{V}_k$ only, (i.e. $\CR_{12}=R_{12}\otimes \mathbb{I}_3$ and so on).
We are interested in $SL(N,\mathbb{C})$ invariant solutions of the YBE.
Therefore, it will be assumed that each space carries a certain representation of
the $SL(N,\mathbb{C})$ group.
For a special choice,
$\mathbb{V}_3=\mathbb{V}_f$, where $\mathbb{V}_f$ is the fundamental
representation of $SL(N,\mathbb{C})$, the YBE turns into $RLL$ relation.
It has the form
\begin{align}\label{}
\mathcal{R}_{12}(u-v)L_1(u)L_2(v)=L_2(v)L_1(u)\mathcal{R}_{12}(u-v)\,,
\end{align}
where $L(u)$ is the Lax operator 
\begin{align}\label{Lax}
L(u)=u+\sum_{mn} e_{mn}\, {E}_{nm}\,.\,&&
\end{align}
With the help of Eq.~(\ref{DE}) the Lax operator~(\ref{Lax}) can be represented as~(see
Eq.~(\ref{EzD}))
\begin{align}\label{}
L(u)=z\,(u-\sigma-D)\,z^{-1}\,.
\end{align}
The form of the operators of right shifts does not depend on a representation, hence
the Lax operator is completely determined by the set of numbers (spectral parameters)
$\{u_1,\ldots,u_N\}$, where
$u_k=u-\sigma_k$, i.e. $L(u)=L(u_1,\ldots,u_N)$.
\vskip 5mm
A solution of the YBE for the principal series representations of the $SL(N,\mathbb{C})$
group, i.e. an $\mathcal{R}$ operator which acts on the tensor product of
the principal series representations $T^{\balpha}\otimes T^{\bbeta}$~\footnote{We accept
the following ``standard'' parametrization for the characters,
${\balpha}(h)=\prod_{k=1}^N
h_{kk}^{-\sigma_k-k}\,\bar
h_{kk}^{-\bar\sigma_k-k}$\,,
$\bbeta(h)=\prod_{k=1}^N
h_{kk}^{-\rho_k-k}\,\bar
h_{kk}^{-\bar\rho_k-k}\,.$ Also, we will use the letters $z$ and $w$ for arguments of the
functions from the representation space of
 $T^{\balpha}\otimes T^{\bbeta}$, $f(z,w)$.
},
has the factorized form~\cite{DM06}
\begin{align}\label{R-factform}
\CR_{12}(u-v)=P_{12}\,\mathbb{R}^{(1)}_{12}(u_1-v_1)
\mathbb{R}^{(2)}_{12}(u_2-v_2)
\ldots
\mathbb{R}^{(N)}_{12}(u_N-v_N)\,.
\end{align}
Here $P_{12}$ is the permutation operator, $P_{12}f(z,w)=f(w,z)$ and the parameters $u_k$,
$v_k$ are defined as follows, $u_k=u-\sigma_k$ and $v_k=v-\rho_k$.
 All operators depend on  holomorphic
and antiholomorphic spectral parameters,
i.e. $\mathbb{R}^{(k)}(\lambda)=\mathbb{R}^{(k)}(\lambda,\bar\lambda)$,
which are subjected to the restriction $\lambda-\bar\lambda\in \mathbb{Z}$, but for brevity we
display the holomorphic variables only.

The defining equation for the operator $\mathbb{R}^{(k)}$ is
\begin{multline}   \label{RkL}
\mathbb{R}^{(m)}_{12}(u_m-v_m)\,
L_1(u_1,\ldots, u_m,\ldots u_N)\,L_2(v_1,\ldots,v_m,\ldots,v_N)=\\
L_1(u_1,\ldots,v_m,\ldots,u_N)\,L_2(v_1,\ldots,u_m,\ldots,v_N)\,\mathbb{R}^{(m)}_{12}(u_m-v_m)\,
\end{multline}
and similar  for the antiholomorphic Lax operators. Thus the operator
$\mathbb{R}^{(m)}_{12}(u_m-v_m)$  exchanges the spectral
parameters $u_m$ and $v_m$ in the Lax operators. It follows from Eq.~(\ref{RkL}) that
$\mathbb{R}^{(m)}(\lambda)$ intertwines the representations $T^{\balpha}\otimes
T^{\bbeta}$ and $T^{\balpha_{m,\lambda}}\otimes T^{\bbeta_{m,-\lambda}}$,
\begin{align}\label{RTT}
\mathbb{R}^{(m)}(\lambda)\,T^{\balpha}(g)\otimes T^{\bbeta}(g)=
T^{\balpha_{m,\lambda}}(g)\otimes T^{\bbeta_{m,-\lambda}}(g)\,\mathbb{R}^{(m)}(\lambda)\,,
\end{align}
where
\begin{align*}\label{}
\balpha_{m,\lambda}(h)=h_{mm}^{-\lambda}\,\bar
h_{mm}^{-\bar\lambda}\,\balpha(h)\,, &&
\bbeta_{m,-\lambda}(h)=h_{mm}^{\lambda}\,\bar h_{mm}^{\bar\lambda}\,\bbeta(h)\,.
\end{align*}
The operator $\mathbb{R}^{(m)}$ is
completely determined by the
``quantum numbers''
of the representations they act on, i.e. by the characters $\balpha$
and $\bbeta$, ($\bsigma$ and $\brho$ )
and the spectral
parameter $\lambda$. We will display the spectral parameter $\lambda$ as an argument of the operator,
$\mathbb{R}^{(m)}(\lambda|\balpha,\bbeta)\to \mathbb{R}^{(m)}(\lambda)$,
and omit the dependence on $\balpha$ and $\bbeta$,
assuming that these parameters are always fixed by a representation the operator acts
on.

\vskip 5mm

The solution of Eq.~(\ref{RkL})
 can be represented
in the form~\cite{DM06}
\begin{multline}\label{RUV}
 \mathbb{R}^{(m)}_{12}(u_m-v_m)=\left(\overleftarrow{\prod_{i=1}^{m-1}} \mathbb{U}_i(u_i-v_m)\right)
\left(\overrightarrow{\prod_{j=m}^{N-1}} \mathbb{V}_j(u_m-v_{j+1})\right)\,\times\\
\mathbb{S}(u_m-v_m)\,
\left(
\overleftarrow{\prod_{j=m}^{N-1}} \mathbb{V}_j(v_{j+1}-v_m)\right)\,
\left(
\overrightarrow{\prod_{i=1}^{m-1}} \mathbb{U}_i(u_m-u_i)\right)\,.
\end{multline}
The operator $\mathbb{S}(\lambda)$ is a multiplication operator
\begin{align}\label{}
\mathbb{S}(\lambda)\,f(z,w)=[(w^{-1}z)_{N1}]^\lambda\,f(z,w)\,,
\end{align}
where
$
[a]^\lambda\equiv a^\lambda \bar a^{\bar\lambda}\,.
$
The function $[a]^\lambda$ is single valued   only if $\lambda-\bar\lambda\in Z$.
This condition is always satisfied in the above construction.  The ordered products are
defined as follows
$
\overrightarrow{\prod_{i=1}^{m}} A_i=A_1\,A_2\ldots A_m
$
and
$
\overleftarrow{\prod_{i=1}^{m}} A_i=A_m\,A_{m-1}\ldots A_1\,.
$

The operators $\mathbb{U}_i(\lambda)\equiv\mathbb{U}_i(\lambda,\bar\lambda)$ and
$\mathbb{V}_i(\lambda)\equiv\mathbb{V}_i(\lambda,\bar\lambda)$ are unitary operators if
$\lambda^*+\bar\lambda=0$. They can be expressed in terms of the operators of right
shifts,
Eq.~(\ref{DD}),
acting on $z$ and $w$ variables, respectively
\begin{align}\label{UV}
\mathbb{U}_i(\lambda,\bar\lambda)=&\left(D_{i+1,i}^{(z)}\right)^\lambda\,\left(\bar
D_{i+1,i}^{(z)}\right)^{\bar\lambda}\,, \\
\mathbb{V}_j(\lambda,\bar\lambda)=&\left(D_{j+1,j}^{(w)}\right)^\lambda\,\left(\bar
D_{j+1,j}^{(w)}\right)^{\bar\lambda}\,.
\end{align}
The operators~(\ref{UV}) are well defined  and can be represented
as  integral operators~\cite{DM06},
\begin{align}\label{prop}
\lbrack \mathbb{U}_i(\lambda)\,\Phi\rbrack(z)=&
A(\lambda)\,\int d^2\zeta\, [\zeta]^{-1-\lambda}\, \Phi(z_\zeta)\,,
\end{align}
where $[\zeta]^{\sigma}=\zeta^{\sigma}\, \bar \zeta^{\bar\sigma}$,
$z_\zeta=z\,\Bigl(1-\zeta\, e_{i+1,i}\Bigr)$ and
 \begin{align}  \label{A-factor}
A(\lambda)\overset{\mathrm{def}}{=}A(\lambda,\bar\lambda)=
\frac1\pi\,
{i^{\bar\lambda-\lambda}}\,
{\Gamma(1+\lambda)}/{\Gamma(-\bar\lambda)}\,.
\end{align}

The operator $\mathbb{U}_k$, with the spectral parameter
$\lambda=\sigma_{k,k+1}=\sigma_k-\sigma_{k+1}$,
 intertwines the representations,
$T^\balpha$ and $T^{\balpha'}$,
\begin{align}\label{UTT}
\mathbb{U}_k(\sigma_{k,k+1})\,T^\balpha(g)=
T^{\balpha'}(g)\,\mathbb{U}_k(\sigma_{k,k+1})\,,
\end{align}
where $\balpha'(h)=\left(h_{kk}/h_{k+1,k+1}\right)^{\sigma_{k,k+1}}
\,\balpha(h)$.

\subsection{Properties of factorizing operators}
The operators $\mathbb{R}^{(k)}$ satisfy a number of remarkable relations~\cite{DM06}
\begin{subequations}\label{propR}
\begin{align}
\label{unity}
\mathbb{R}^{(m)}_{12}(0) &= \II\,,
\\[2mm]
\label{p1}
\mathbb{R}^{(m)}_{12}(\lambda)\mathbb{R}^{(m)}_{12}(\mu) &= \mathbb{R}^{(m)}_{12}(\lambda+\mu)\,,
\\[2mm]
\label{R-R}
\mathbb{R}_{12}^{(m)}(\lambda)\,\mathbb{R}_{23}^{(n)}(\mu) &=
\mathbb{R}_{23}^{(n)}(\mu)\mathbb{R}_{12}^{(m)}(\lambda)\,, \qquad\qquad
(n\neq m)\,,\\[2mm]
\label{R-R-R}
\mathbb{R}_{12}^{(m)}(\lambda)\,\mathbb{R}_{23}^{(m)}(\lambda+\mu)\,\mathbb{R}_{12}^{(m)}(\mu)
 &=
\mathbb{R}_{23}^{(m)}(\mu)\,\mathbb{R}_{12}^{(m)}(\lambda+\mu)\,\mathbb{R}_{23}^{(m)}(\lambda)\,,\\[2mm]
\label{R-2-R}
\mathbb{R}^{(m)}_{12}(\lambda-\sigma_m+\rho_m)\,\mathbb{R}^{(n)}_{12}(\lambda-\sigma_n+\rho_n) &=
\mathbb{R}^{(n)}_{12}(\lambda-\sigma_n+\rho_n)\,\mathbb{R}^{(m)}_{12}(\lambda-\sigma_m+\rho_m)\,.
\end{align}
\end{subequations}
These relations are sufficient to prove that the $\mathcal{R}-$operator~(\ref{R-factform})
satisfy the YBE~\cite{DM06}.
Next, taking into account Eqs.~(\ref{DD}),~(\ref{UV}) and (\ref{RUV}) one deduces the following
commutation relations for the operators $\mathbb{R}^{(m)}$
\begin{subequations}\label{Rz}
\begin{align}
\mathbb{R}^{(m)}(\lambda)\, (z^{-1})_{kj}&=(z^{-1})_{kj}\,\mathbb{R}^{(m)}(\lambda)\,,
&&\text{for}\quad k> m\,,\\
\mathbb{R}^{(m)}(\lambda)\,w_{kj}&=w_{kj}\,\mathbb{R}^{(m)}(\lambda)\,,
&&\text{for}\quad  j< m\,
\end{align}
\end{subequations}
and
\begin{subequations}\label{DR}
\begin{align}\label{}
D_{k+1,k}^{(z)} \,\mathbb{R}^{(m)}(\lambda)&=\mathbb{R}^{(m)}(\lambda)\,D_{k+1,k}^{(z)},
&&\text{for}\quad k> m+1\,,\\
D_{k+1,k}^{(w)}\, \mathbb{R}^{(m)}(\lambda)&=\mathbb{R}^{(m)}(\lambda)\,D_{k+1,k}^{(w)},
&&\text{for}\quad k< m-1\,.
\end{align}
\end{subequations}
Finally, one can see from the representation~(\ref{RUV}) that the operator $\mathbb{R}^{(m)}$
depends only on the spectral parameters,
$u_1,\ldots,u_m$ and $v_m,\ldots,v_N$. Namely, it depends on the spectral
parameter $\lambda=u_m-v_m$ and
\begin{align}\label{uv-dependence}
u_i-u_m=\sigma_m-\sigma_i,\quad  i<m,\quad\text{and}\quad v_m-v_j=\rho_j-\rho_m,\quad j>m.
\end{align}
It means that
\begin{align}
\mathbb{R}^{(m)}(\lambda|\balpha,\bbeta)=\mathbb{R}^{(m)}(\lambda|\balpha',\bbeta')
\end{align}
if the characters satisfy the following relations
\begin{align*}
\frac{\balpha(h)}{\balpha'(h)}=f(\Delta_m(h),\ldots,\Delta_{N-1}(h))&&\text{and}&&
\frac{\bbeta(h)}{\bbeta'(h)}=\varphi(\Delta_1(h),\ldots,\Delta_{m-1}(h))\,.
\end{align*}
That is the ratio ${\balpha(h)}/{\balpha'(h)}$ does not depend on $\Delta_k$, $k=1,\ldots,m-1$
and similarly for  $\bbeta$.

\subsection{Factorizing operators in the coherent states basis}
Let us calculate the action  of the operator $\mathbb{R}^{(m)}$ in the coherent state
basis
\begin{align}\label{cDD}
\Delta^{\bsigma\brho}(z,w|\alpha,\beta)\equiv
\Delta^\bsigma(z,\alpha)\,\Delta^{\brho}(w,\beta)\,,
\end{align}
where $z,w\in Z_-$ and $\alpha,\beta\in Z_+$.
We will use the following notations
\begin{align}\label{alphaz}
\alpha\,z=z_{\alpha}\, d_{z,\alpha}\, \alpha_z\,, &&
\beta\,w=w_{\beta}\, d_{w,\beta}\, \beta_w\,,
\end{align}
where $d_{z,\alpha},d_{w,\beta}$ are diagonal matrices,
$z_{\alpha}$, $w_{\beta}\in Z_-$ and
$ \alpha_z, \beta_w\in Z_+ $.
{\lemma \label{R-coherent}
The action of the operator $\mathbb{R}^{(m)}_{12}(\lambda)$ on the state~(\ref{cDD})
is given by
\begin{align}\label{R-act}
\Big[\mathbb{R}^{(m)}_{12}(\lambda)\,\Delta^{\bsigma\brho}\Big](z,w|\alpha,\beta)=
f^{(m)}_{\bsigma\brho}(\lambda)\,
\left[\left(\beta_w\,w^{-1}\, z\, \alpha_z^{-1} \right)_{mm}\right]^\lambda\,
\Delta^{\bsigma\brho}(z,w|\alpha,\beta)\,,
\end{align}
where $[a]^\lambda\equiv a^\lambda \bar a^{\bar\lambda}$.
The  prefactor $f^{(m)}_{\bsigma\brho}(\lambda)$ is 
\begin{multline}\label{f-def}
f^{(m)}_{\bsigma\brho}(\lambda)=
\prod_{k=1}^{m-1}(-1)^{\lambda-\bar\lambda}\frac{A(\lambda-\sigma_{km})}{A(-\sigma_{km})}\,
\prod_{j=m+1}^N\frac{A(\lambda-\rho_{mj})}{A(-\rho_{mj})}=\\
=\prod_{k=1}^{m-1}(-1)^{\lambda-\bar\lambda}\frac{A(u_k-v_m)}{A(u_k-u_m)}\,
\prod_{j=m+1}^N\frac{A(u_m-v_j)}{A(v_m-v_j)}\,,
\end{multline}
where
$$
A(\lambda)=\frac1\pi\,
{i^{\bar\lambda-\lambda}}\,
{\Gamma(1+\lambda)}/{\Gamma(-\bar\lambda)}\,
$$
and
\begin{align}\label{rel-uv}
\lambda=u_m-v_m\,, &&u_m=u-\sigma_k\,, &&v_k=v-\rho_k\,.
\end{align}}
%
The proof of the Lemma is rather technical and  can be  found in Appendix~\ref{lemma1}.
Now we want to discuss Eq.~(\ref{R-act}) in more details. First, we notice that the
r.h.s. of  Eq.~(\ref{R-act}) is given (up to the prefactor $f^{(m)}_{\bsigma\brho}(\lambda)$)
by the product of two functions
\begin{align}\label{Ku}
\mathcal{K}^m_{\lambda,\bsigma,\brho}(z,w|\alpha,\beta)&=
\left(\beta_w\,w^{-1}\, z\, \alpha_z^{-1} \right)_{mm}^\lambda\,
\left(\prod_{k=1}^{N-1}(\Delta_{k}(\alpha z))^{\sigma_{k,k+1}-1}
(\Delta_{k}(\beta w))^{\rho_{k,k+1}-1} \right)\,\\
\intertext{and}
\label{bKu}
\bar{\mathcal{K}}^m_{\bar\lambda,\bar\bsigma,\bar\brho}(z, w|\alpha,\beta)&=
\left(\left(\beta_w\,w^{-1}\, z\, \alpha_z^{-1} \right)^*_{mm}\right)^{\bar\lambda}\,
\left(\prod_{k=1}^{N-1}(\Delta_{k}(z^\dagger\alpha^\dagger))^{\bar\sigma_{k,k+1}-1}
(\Delta_{k}(w^\dagger\beta^\dagger))^{\bar\rho_{k,k+1}-1} \right)\,.
\end{align}
The function $\mathcal{K}\,(\bar{\mathcal{K}})$ is an (anti)holomorphic function of
$z,w,\alpha,\beta$ in the vicinity of the point $z=w=\alpha=\beta=1$.

Further, it follows from  Eq.~(\ref{RkL}) that the function
$\mathcal{K}^m_{\lambda,\bsigma,\brho}(z,w|\alpha,\beta)$ satisfies the following equation
\begin{multline}   \label{RKL}
\widetilde L_1^{\alpha}(u_1,\ldots, u_m,\ldots u_N)\,
\widetilde L_2^{\beta}(v_1,\ldots,v_m,\ldots,v_N)\,
\mathcal{K}^m_{u_m-v_m,\bsigma,\brho}(z,w|\alpha,\beta)
=\\
L_1^{z}(u_1,\ldots,v_m,\ldots,u_N)\,L_2^{w}(v_1,\ldots,u_m,\ldots,v_N)\,
\mathcal{K}^m_{u_m-v_m,\bsigma,\brho}(z,w|\alpha,\beta)\,.
\end{multline}
Here the Lax operators $\widetilde L_1 (\widetilde L_2)$
are  given by
\begin{align}\label{tilde-Lax}
\widetilde L(u)=u-\sum_{mn} e_{mn}\, \widetilde {E}_{nm}\,,
\end{align}
where the generators $\widetilde {E}_{nm}$
correspond to the representation $\widetilde T^{-\bsigma} (\widetilde T^{-\brho})$,
see Eqs.~(\ref{TtT}),~(\ref{Ttilde}). The superscript of
the Lax operator ($L^z$,~$\widetilde L^{\alpha}$) indicates the variable it acts on.
Equation~(\ref{RKL}) follows directly from 
Eqs.~(\ref{EwE}),~(\ref{RkL}),~(\ref{R-act}).
In the next section we show that the
function
$\mathcal{K}^m_{\lambda,\bsigma,\brho}(z,w|\alpha,\beta)$ defines a factorizing
operator on the tensor product of Verma modules.

\section{${{sl}}(N)$ invariant $\mathcal{R}$ operator for generic highest weight representations}
\label{roperator}
In this section we construct  $sl(N)$ invariant solution of the YBE  on the tensor
product of two generic highest weight representations of the $sl(N)$ Lie algebra (Verma
modules).

Let $\mathbb{V}$ be a linear space of polynomials of  arbitrary degree in $z_{ki}$,
\begin{align}\label{Vm}
\mathbb{V}=\Big\{P(z_{21},z_{31},\ldots,
z_{NN-1}),\quad \deg(P)<\infty \Big \}.
\end{align}
The homomorphism
\begin{align}
\pi^\bsigma: \mathcal{E}_{ki}\to E_{ki}=
-\sum_{m\leq n}z_{im}\,\Bigl(D_{nm}+\delta_{nm}\,\sigma_m\Bigr)\,(z^{-1})_{nk}\,
\end{align}
(see Eq.~(\ref{EDs}) for a definition of $D_{nm}$) defines
a representation of the $sl(N)$ algebra on the space~$\mathbb{V}$.
The operators $E_{ki}$ are completely determined by the parameters
$\{\sigma_1,\ldots,\sigma_N\}$.
More precisely, they depend on the differences $\sigma_{nm}=\sigma_n-\sigma_m$. In
order to stress  a similarity with the $SL(N,\mathbb{C})$ case, it is convenient to
specify the representation by  a function
$\balpha(h)=\prod_{k=1}^N h_{kk}^{-k-\sigma_k}$\,,
where $h$ is an upper triangular matrix with unit determinant.  Henceforth, we will use  both
notations,
$\pi^\balpha$ and $\pi^\bsigma$, for a representation of the  $sl(N)$ algebra.
A representation $\pi^\bsigma$
is irreducible if none of the differences $\sigma_{ik}=\sigma_i-\sigma_k$, $i<k$, is a positive
integer~\cite{Verma,BGG}. We will assume that this condition is  fulfilled.

The highest weight vector, $\upsilon_0$, ($E_{ik}\upsilon_0=0$, for $i>k$) of the Verma
module~(\ref{Vm}) is  given by $\upsilon_0=1$.  
For the highest weight
$\boldsymbol{\lambda}=(\lambda_1,\ldots,\lambda_{N-1})$,
($(E_{kk}-E_{k+1,k+1})\cdot \upsilon_0\equiv\lambda_k\upsilon_0$)
one finds $\lambda_k=\sigma_{k+1}-\sigma_k+1$. 
In  case if all components of the weight vector $\blambda$ are negative integer, $\lambda_k=-n_k$,
$n_k\geq 0$, $k=1,\ldots,N-1$ the Verma module has a finite dimensional invariant subspace.
This subspace is a finite dimensional representation of $sl(N)$ algebra
which corresponds to the Young tableau specified by the partition 
$\{\ell_1,\ldots,\ell_{N-1}\}$, where $\ell_k=\sum_{i=k}^{N-1}n_k$ is the length of the
$k-$th row in the Young tableau.
 
\vskip 5mm

\subsection{Bilinear form and kernel of an operator}
We define the following linear combinations of the Cartan generators
$E_{kk}$ and the unit operator,
\begin{align}\label{H-p}
H_p=&\sum_{k=1}^p (E_{kk}+\sigma_k+k-N)=\sum_{m=p+1}^N \sum_{k=1}^p
z_{mk}\frac{\partial}{\partial z_{mk}}=
\sum_{m=p+1}^N \sum_{k=1}^p \tilde z_{mk}\frac{\partial}{\partial \tilde z_{mk}}\,,
\end{align}
where $p=1,\ldots, N-1$ and $\tilde z=z^{-1}$. The space $\mathbb{V}$ is a direct sum of
the weight subspaces $\mathbb{V}_h$,
\begin{align}\label{V-direct}
\mathbb{V}=\sum_{h\in \mathbb{Z}_+^{N-1}} \oplus\mathbb{V}_h\,,\qquad
\mathbb{V}_h=\{v\in \mathbb{V}|(H_p-h_p)v=0,p=1,\ldots,N-1\}\,.
\end{align}
Each subspace $\mathbb{V}_h$ has  a finite dimension. The union of the bases in all
$\mathbb{V}_h$ gives a basis in~$\mathbb{V}$. We will mostly use the following basis
\begin{align}\label{enz}
e_n(z)=\prod_{i>k} {z_{ik}^{n_{ik}}}\,,
\end{align}
where $n$ is a multi-index, $n=\{n_{21},\ldots,n_{NN-1}\}$.

Let $\widebar{\mathbb{V}}$ be a linear space of polynomials of arbitrary degree in 
$\bar z_{ki}=z_{ki}^*$, 
\begin{align}
\widebar{\mathbb{V}}=\Big\{P(\bar z_{21},\bar z_{31},\ldots,
\bar z_{NN-1}), \quad \deg(P)<\infty \Big \}
\end{align}
and  $\varphi$ be an antilinear map $\mathbb{V}\to \widebar{\mathbb{V}}$ defined by
$\varphi(e_n)=\bar e_n=\prod_{i>k} {\bar z_{ik}^{n_{ik}}}$. We put
$\widebar{\mathbb{V}}_h=\varphi(\mathbb{V}_h)$.

Let  $\Omega$ be a bilinear form  on the product
$\widebar{\mathbb{V}}\times\mathbb{V}$ such that
\begin{subequations}
\label{Omega}
\begin{align}
& \Omega(\bar v, u)=0,\quad \text{if}\quad \bar v\in\widebar{\mathbb{V}}_h\quad \text{and}\quad
u\in \mathbb{V}_{h'},\quad
h \neq h'\\
&\text{ the matrix}\quad \Omega_{nm}=\Omega(\bar e_n, e_m)\quad \text{is invertible}.
\end{align}
\end{subequations}
Let  $\mathbb{A}$ be a
linear operator on the space $\mathbb{V}$
and   $A_{nm}$ be
its matrix in the basis $e_{n}$, $\mathbb{A}e_n=\sum_m e_m A_{mn}$.
We will refer to a function
\begin{align}\label{}
\mathcal{A}(z,w)=\sum_{nm} e_n(z) (A\Omega^{-1})_{nm} \overline{e_m(w)}\,
\end{align}
as a kernel of the operator $\mathbb{A}$.
The kernel $\mathcal{A}(z,w)$ is  a (anti)holomorphic function  in $z(w)$
in the  vicinity of the point $z=w=1$ ($z_{ik}=w_{ik}=0$)
on condition that  the series converges.
An action of the operator $\mathbb{A}$ on an arbitrary vector
from $\mathbb{V}$, (which is a polynomial in $z$) can be represented as
\begin{align}\label{}
[\mathbb{A}\,P](z)=\Omega({\mathcal{A}(z,w)},P(w))\,.
\end{align}
The bilinear form $\Omega$ is completely determined by a kernel of the unit operator
(reproducing kernel)
\begin{align}\label{}
\mathcal{I}(z,w)=&\sum_{nm} e_n(z)\, \Omega^{-1}_{nm} \,\overline{e_m(w)}\,. 
\end{align}
It is clear that the kernel of an operator does not depend on  a choice of the 
basis. In particularly, the kernel $A(z,w)$ of the operator $\mathbb{A}$ can be obtained as
\begin{align}\label{}
\mathcal{A}(z,w)=\mathbb{A}\,\mathcal{I}(z,w)\,.
\end{align}

Let $\pi^\bsigma$ be an irreducible representation of the 
$sl(N)$ algebra on the vector space $\mathbb{V}$.
Henceforth  we will assume that  the space $\mathbb{V}$ is equipped with a bilinear form
$\Omega_\bsigma$ such that the reproducing kernel has the form%
\begin{align}\label{I-rep}
\mathcal{I}^\bsigma(z,w)=\prod_{k=1}^{N-1}(\Delta_{k}(w^\dagger z))^{\sigma_{k,k+1}-1}\,.
\end{align}
It is easy to derive from  Eqs.~(\ref{Tg}) and (\ref{Ttilde}) that the kernel
$\mathcal{I}^\bsigma(z,w)$ satisfies the equation
\begin{align}\label{EEt}
E_{ki}^{(z)}\,\mathcal{I}^\bsigma(z,w)
=-\widetilde E_{ki}^{(\bar w)}\,\mathcal{I}^\bsigma(z,w)\,,
\end{align}
where   $E_{ik}=\pi^\bsigma(e_{ik})$ and
$\widetilde E_{ki}=\tilde\pi^{-\bsigma}(e_{ik})$ are the holomorphic generators corresponding to the
representations $T^\bsigma$  (Eq.~(\ref{Tg})) and $\widetilde T^{-\bsigma}$
(Eqs.(\ref{TtT}),(\ref{Ttilde}))~\footnote{
The representation $\widetilde T^{-\bsigma}$ was defined on  functions on the group $Z_+$,
$f(\alpha)$, so that $\widetilde E_{ki}^{(\bar w)}=\widetilde
E_{ki}(\alpha)|_{\alpha=w^\dagger}$ }.
It follows from  Eq.~(\ref{EEt}) that the form $\Omega_\bsigma$
 (for the irreducible representation) satisfies the properties~(\ref{Omega}).
From  Eq.~(\ref{EEt}) one derives
\begin{align}\label{EtE}
\Omega_\bsigma(\widetilde E_{ki}{Q},P)+\Omega_\bsigma({Q},E_{ki}P)=0\,.
\end{align}
As usual, a bilinear form on the tensor product of two (or more) representations
$\pi^\bsigma\otimes \pi^\brho$,
is defined as
$$
\Omega_{\bsigma\brho}(Q_1\otimes Q_2, P_1\otimes P_2)=\Omega_\bsigma(Q_1,P_1)\cdot
\Omega_\brho(Q_2,P_2).
$$

It will be useful to have  a more functional definition for the bilinear form $\Omega_\bsigma$.
Let us put for any two polynomials $P(z)$ and $Q(\bar z)$
\begin{align}\label{def-mu}
B_\bsigma(Q, P)=c_N(\bsigma)\int Dz \,\mu_\bsigma(z) \,Q(\bar z)\, P(z)\,,
\end{align}
where
\begin{align}\label{mu-z}
\mu_\bsigma(z)=\prod_{k=1}^{N-1}(\Delta_{k}(z^\dagger z))^{-\sigma_{k,k+1}-1}\,.
\end{align}
Since $P$ and $Q$ are polynomials, the integral in~(\ref{def-mu}) converges in some
region of $\sigma_{k,k+1}$ and, as will be shown later,
defines a meromorphic function of $\sigma_{ik}$, which
we take for a definition of the l.h.s of~(\ref{def-mu}) for arbitrary $\sigma_{ik}$.
Note, that for  positive integer $\sigma_{k,k+1}$ the integral~(\ref{def-mu}) defines the
invariant $SU(N)$ scalar product.
It is easy to check that the bilinear form~(\ref{def-mu}) with the measure (\ref{mu-z})
results in Eq.~(\ref{EtE}). Hence the bilinear form~$B_\bsigma$, Eq.~(\ref{def-mu}) and
the form $\Omega_\bsigma$  coincide up to a prefactor.
The two forms coincide identically at the following normalization
\begin{align}\label{}
 c_N(\bsigma)=\pi^{-{N(N-1)}/{2}}\left(\prod_{1\leq i<k\leq N}\sigma_{ik}\right)
 \end{align}
($B_\bsigma(Q,P)=1=\Omega_{\bsigma}(Q,P)$ for $Q(\bar z)=P(z)=1$).
For example, for $N=2$ Eq.~(\ref{def-mu}) becomes
\begin{align}\label{}
\Omega_\bsigma(Q, P)=\frac{2j+1}{\pi}\int d^2z \,\frac{Q(\bar z)\, P(z)}{(1+z\bar z)^{2j+2}}\,,
\end{align}
where $2j=\sigma_{12}-1$.

It is useful to extend the space $\mathbb{V}$ in order to include into  consideration
non-polynomial functions which are analytic in the  vicinity of the point 
$z=1$ $ (z_{ik}=0)$. This allows us
to consider  finite
transformations of functions from $\mathbb{V}$, $P(z)\to f(z)=\alpha(h)\,P(z\bar g)$. The
function $f(z)$ is no more a polynomial,
however, if  $g$  is sufficiently close  to unity %
then $f(z)$  is an
analytic function of $z$ in the vicinity
of the point $z=1$ ( $f(z)=\sum_{n}c_n e_n(z)=\sum_h f_h(z)$, where $f_h$ is a projection of the
vector $f$ to the weight subspace $\mathbb{V}_h$ ).
 For such functions the bilinear form is defined as the sum,
$\Omega_\bsigma(\bar \psi,f)=\sum_h \Omega_{\bsigma}(\bar\psi_h,f_h)$, 
in  case if  the series converges.

\subsection{$\mathcal{R}-$matrix  and factorizing operators.}
Our immediate purpose  in this subsection is to construct the factorizing operators
$\mathbb{R}^{(m)}$, $m=1,\ldots,N$ which act on the tensor product of two Verma modules
$\mathbb{V}\otimes \mathbb{V}$
and solve the $RLL$ relation
\begin{multline}   \label{R-L}
\mathbb{R}^{(m)}_{12}(u_m-v_m)\,
L_1(u_1,\ldots, u_m,\ldots u_N)\,L_2(v_1,\ldots,v_m,\ldots,v_N)=\\
L_1(u_1,\ldots,v_m,\ldots,u_N)\,L_2(v_1,\ldots,u_m,\ldots,v_N)\,\mathbb{R}^{(m)}_{12}(u_m-v_m)\,.
\end{multline}
The parameters $u_m,v_m$ are defined by Eq.~(\ref{rel-uv}).
It is straightforward to check that the operator $\mathbb{R}^{(m)}_{12}(\lambda)$
intertwines the representations $\pi^\balpha\otimes \pi^\bbeta$
($\equiv \pi^\bsigma\otimes \pi^\brho$) and
$\pi^{\balpha_{m,\lambda}}\otimes \pi^{\bbeta_{m,-\lambda}}$
($\equiv \pi^{\bsigma'}\otimes \pi^{\brho'}$),
\begin{align}\label{twineR}
\mathbb{R}^{(m)}_{12}(\lambda)\,\pi^\balpha\otimes \pi^\bbeta=
\pi^{\balpha_{m,\lambda}}\otimes \pi^{\bbeta_{m,-\lambda}}\,
\mathbb{R}^{(m)}_{12}(\lambda)\,,
\end{align}
where $\balpha_{m,\lambda}(h)=h_{mm}^{-\lambda}\,\balpha(h)$ and
$\bbeta_{m,-\lambda}(h)=h_{mm}^{\lambda}\,\bbeta(h)$.

Let $\mathcal{R}^{(m)}_{\lambda}(z,w|\bar\alpha,\bar\beta)$ be a kernel of the operator
$\mathbb{R}^{(m)}_{12}$,
$$[\mathbb{R}^{(m)}_{12}\psi](z,w)=
\Omega_{\bsigma'\brho'}(\mathcal{R}^{(m)}_{\lambda}(z,w|\bar\alpha,\bar\beta),\psi(\alpha,\beta)).
$$
 Using
property~(\ref{EtE}) one  easily derives the defining equation for
$\mathcal{R}^{(m)}_{\lambda}(z,w|\bar\alpha,\bar\beta)$
\begin{multline}\label{newRKL}
\widetilde L_1^{\bar \alpha}(u_1,\ldots, u_m,\ldots u_N)\,
\widetilde L_2^{\bar\beta}(v_1,\ldots,v_m,\ldots,v_N)\,
\mathcal{R}^{(m)}_{\lambda}(z,w|\bar\alpha,\bar\beta)
=\\
L_1^{z}(u_1,\ldots,v_m,\ldots,u_N)\,L_2^{w}(v_1,\ldots,u_m,\ldots,v_N)\,
\mathcal{R}^{(m)}_{\lambda}(z,w|\bar\alpha,\bar\beta)\,,
\end{multline}
where $\lambda=u_m-v_m$.
We recall here that
$$
\widetilde L_1(u_i)=u-\sum_{nm} e_{nm}\,\widetilde E^{(1)}_{mn}
$$
and similarly for $\widetilde L_2$.  Let us notice that  Eq.~(\ref{newRKL})
coincides identically with Eq.~(\ref{RKL}) whose solution is given by the
function $\mathcal{K}^m_{\lambda,\bsigma,\brho}(z,w|\bar\alpha,\bar\beta)$, Eq.~(\ref{Ku}).
Since the function $\mathcal{K}^m_{\lambda,\bsigma,\brho}(z,w|\bar\alpha,\bar\beta)$ is an
analytic function of $z,w,\bar\alpha,\bar\beta$ in the vicinity of the point
$z=w=\alpha=\beta=1$ ($z_{ik}=\ldots=\beta_{ik}=0$) it defines an operator on
the tensor product of Verma modules
$\mathbb{V}\otimes\mathbb{V}$. Thus, we have proven the following statement.

{\lemma The operator $\mathbb{R}^{(m)}_{12}(\lambda)$:
$\mathbb{V}\times \mathbb{V}\to \mathbb{V}\times \mathbb{V}$
defined by the kernel
\begin{align}
\mathcal{R}^{(m)}_{\lambda,\bsigma\brho}(z,w|\bar\alpha,\bar\beta)
=A_m\,\left(\bar\beta_w\,w^{-1}\, z\, \bar\alpha_z^{-1}
\right)_{mm}^\lambda\,
\mathcal{I}^\bsigma(z,\alpha)\,\mathcal{I}^\brho(w,\beta)\,,
\label{r-kernel}
\end{align}
where $A_m$ is some constant, solves $RLL$ relation~(\ref{R-L}).}
\vskip 3mm

 As follows from~Eq.~(\ref{r-kernel})
the operator $\mathbb{R}^{(m)}$  depends both on the parameters  $\bsigma$
and $\brho$ and  the spectral parameter~$\lambda$. We will display explicitly the dependence
on a spectral parameter only. It means that  a product of two operators, for instance
$\mathbb{R}^{(m)}_{12}(\mu)\,\mathbb{R}^{(m)}_{12}(\lambda)$, written in an explicit form, turns
into $\mathbb{R}^{(m)}_{12}(\mu|\bsigma',\brho')\,\mathbb{R}^{(m)}_{12}(\lambda|\bsigma,\brho)$,
where the parameters $\bsigma',\brho'$ are determined by Eq.~(\ref{twineR}).

It is easy to see from Eq.~(\ref{r-kernel}) that  matrix elements of the operator
$\mathbb{R}^{(m)}_{12}(\lambda)$ (modulo the prefactor $A_m$)
are  analytic  functions of the spectral parameter
$\lambda$. Moreover, they are meromorphic functions in $\bsigma$ and $\brho$, 
which specify  representations of
the $sl(N)$ algebra on the space $\mathbb{V}\otimes\mathbb{V}$.   The position of the poles
 corresponds to the points of reducibility
of the representations $\pi^{\bsigma}$ and $\pi^{\brho}$.
\vskip 2mm

The normalization factor $A_m$ in~(\ref{r-kernel}) is, in general, an arbitrary function of
 $\lambda$, $\bsigma$, $\brho$: $A_m=A_m(\lambda,\bsigma,\brho)$.
However, the operators $\mathbb{R}^{(m)}_{ik}$  satisfy the relations~(\ref{propR})
only if  the functions $A_m$ obey some restrictions.
They can be easily read off Eqs.~(\ref{propR}), for instance,
\begin{align}\label{}
A_m(0,\bsigma,\brho)=1\,,\qquad A_m(\mu,\bsigma',\brho')A_m(\lambda,\bsigma,\brho)=
A_m(\mu+\lambda,\bsigma,\brho)\,
\end{align}
and so on. The simplest normalization which satisfy all the requirements is
$A_m=1$. Choosing another, $SL(N,\mathbb{C})$ induced normalization,
one can get rid of unessential prefactors in some formulae for  transfer matrices.
In this normalization the factor $A_m$ reads
\begin{align}\label{AslN}
A_m(u-v,\bsigma,\brho)=f_m(u-v)\prod_{k=1}^{m-1}\frac{\Gamma(u_k-v_m+1)}{\Gamma(u_k-u_m+1)}\,
\prod_{j=m+1}^N\frac{\Gamma(u_m-v_j+1)}{\Gamma(v_m-v_j+1)}\,,
\end{align}
where
$ f_m(\lambda)=1$ for even $m$ and
$f_m(\lambda)=e^{i\pi\lambda}$ for odd $m$.
We will assume that the normalization factor
possesses all necessary properties.
Its explicit form will be irrelevant for  further discussion.
\vskip 5mm

Let us  prove that the operators $\mathbb{R}^{(m)}_{ik}(\lambda)$
satisfy the relations~(\ref{propR}).  The first of them,
$\mathbb{R}^{(m)}_{ik}(0)=\mathbb{I}$, follows directly from Eq.~(\ref{r-kernel}).
The second one,~(\ref{p1}),
requires a special analysis and will be discussed in Appendix~\ref{AppB}.

Going on to the proof of the relations~(\ref{R-R}),~(\ref{R-R-R})~(\ref{R-2-R}) we notice
that the product of the operators $\mathbb{R}^{(m)}$ in the l.h.s and r.h.s of the
corresponding equations
results in the same permutations of the spectral parameters $u_k,v_k, w_k$ in
the product of two (three) Lax operators. Let us show  that two operators which
result in the same permutation of the spectral parameters in a product of Lax
operators coincide up to a normalization. Namely, if
\begin{align}\label{a-b}
A(u-v)\,L_1(u)\,L_2(v)=&L'_1(u')\,L'_2(v')\,A(u-v)\,,\notag\\
 B(u-v)\,L_1(u)\,L_2(v)=&L'_1(u')\,L'_2(v')\,B(u-v)\,,
\end{align}
then $A(\lambda)=c B(\lambda)$.  It follows from Eqs.~(\ref{a-b}) that the operator
$C(\lambda)=B^{-1}(\lambda)\,A(\lambda)$  commutes with the product of Lax operators
\begin{align}\label{}
C(u-v)\,L_1(u)\,L_2(v)=L_1(u)\,L_2(v)\,C(u-v)\,.
\end{align}
We will assume that the operators $A(\lambda)$ and $B(\lambda)$ and, hence the operator $C(\lambda)$,
are analytic operators that means that their matrix elements are analytic
(meromorphic) functions of $\lambda$. The following lemma states that an operator with
such properties is a multiple of the unit operator.

{\lemma \label{irrec} Let $\pi^{(1)}$ and
$\pi^{(2)}$ be irreducible highest weight representations of the $sl(N)$
algebra on  the spaces $\mathbb{V}_1$ and $\mathbb{V}_2$.
If an analytic operator $A(\lambda): \mathbb{V}_1\otimes\mathbb{V}_2\to\mathbb{V}_1\otimes\mathbb{V}_2 $
commutes with a product of Lax operators
\begin{align}\label{ALL}
A(u-v)\,L_1(u)\, L_2(v)=L_1(u)\,L_2(v)\, A(u-v)\,,
\end{align}
then $A(\lambda)=a(\lambda)\,\mathbb{I}$.
}
\begin{proof}
It follows from (\ref{ALL}) that
\begin{align}\label{AE}
[A(\lambda), E_{ki}^{(1)}(\lambda)]=[A(\lambda), E_{ki}^{(2)}(\lambda)]=0\,,
\end{align}
where
\begin{align}\label{def_E}
E_{ki}^{(1)}(\lambda)=E_{ki}^{(1)}-\frac{1}{\lambda}\sum_m E_{km}^{(2)}E_{mi}^{(1)}\,,&&
E_{ki}^{(2)}(\lambda)=E_{ki}^{(2)}+\frac{1}{\lambda}\sum_m E_{km}^{(2)}E_{mi}^{(1)}\,.
\end{align}
The space $\mathbb{V}\otimes\mathbb{V}$ can be decomposed into the direct
sum of invariant subspaces of the Cartan generators, $\mathbb{V}\otimes\mathbb{V}=\sum_{h}
\mathbb{V}_h$, $\dim \mathbb{V}_h=N_h<\infty$.
Since $[E^{(1)}_{kk}+E^{(2)}_{kk},
A(u)]=0$, the subspace $\mathbb{V}_h$ is an invariant subspace of the operator $A(u)$.
Let $e_n^h$, $n=1,\ldots, N_h$ be a basis in the subspace~$\mathbb{V}_h$.
Since the representations $\pi^{(1)}$, $\pi^{(2)}$ are irreducible, the basis vectors
$e_n^h$ are given by  linear combinations of the vectors 
$$
\prod_{k>i} (E_{ki}^{(1)})^{n_{ki}} \prod_{j>m} (E_{jm}^{(2)})^{n_{jm}}\upsilon_0,
$$
 where
$\upsilon_0$ is the highest weight vector, $\upsilon_0=1$. So we write $e_n^h=e_n^h(E^{(1)},E^{(2)})$.
Now let us consider a set of the vectors $e_n^h(\lambda)= e_n^h(E^{(1)}(\lambda),E^{(2)}(\lambda))$.
It follows from Eq.~(\ref{def_E}) that for a sufficiently large~$\lambda$, the vectors
$e_n^h(\lambda)=e_n^h+\mathcal{O}(1/\lambda)$,
$n=1,\ldots,N_h$ are linearly independent. Hence they  form a basis in
the subspace~$\mathbb{V}_h$. By virtue of Eq.~(\ref{AE}) one finds that
$A(\lambda) e_n^h(\lambda)=a(\lambda)e_n^h(\lambda)$, where $a(\lambda)=A(\lambda)\upsilon_0$.
Thus, we have proven that
\begin{align}\label{Aa}
A(\lambda)=a(\lambda)\, \mathbb{I}
\end{align}
on an arbitrary subspace $\mathbb{V}_h$  in some region of
$\lambda$. Due to assumed analyticity Eq.~(\ref{Aa}) is valid for an arbitrary $\lambda$.
Therefore Eq.~(\ref{Aa}) holds for the whole space $\mathbb{V}_1\otimes\mathbb{V}_2$.
\end{proof}
\vskip 5mm

It is clear that the proof of  {\sl lemma~\ref{irrec}}
can be easily extended to the case of an arbitrary number
of Lax operators. Namely, if an analytic operator $A(\lambda_1,\lambda_2,\lambda_{M-1})$ satisfy
the equation
\begin{align}\label{}
A(\lambda_i)\,L_1(u)L_2(u+\lambda_1)\ldots L_M(u+\lambda_{M-1})=
L_1(u)L_2(u+\lambda_1)\ldots L_M(u+\lambda_{M-1})\,A(\lambda_i)\,,
\end{align}
then $A(\lambda_i)\sim \mathbb{I}$. As was explained earlier this
result implies that the l.h.s and r.h.s of
Eqs.~(\ref{R-R}),~(\ref{R-R-R}) and (\ref{R-2-R}) are equal to
each other up to some factor, $\varkappa$. It can be easily
checked by examining the action of the operators on the highest
weight vector, $\upsilon_0=1$, that  $\varkappa=1$. We also notice
here that {\sl Lemma~\ref{irrec}} implies the uniqueness of  the
solution of the $RLL$ relation~(\ref{R-L}).

\vskip 5mm

The  commutation relations~(\ref{p1}),~(\ref{DR}) for the operators $\mathbb{R}^{(m)}$ 
are vital for our analysis of  transfer matrices, see Sect.~\ref{TmQ}.
The proof of these relations makes use of the explicit form of the
factorizing operators,~Eq.~(\ref{r-kernel}), and the invariance property of the
bilinear form~$\Omega_\bsigma$.
First of all, we prove the following lemma:
{\lemma Let $g\in SL(N)$ and $\Delta_k(g)\neq 0$, $k=1,\ldots,N-1$. Then
\begin{align}\label{tpq}
\Omega_\bsigma(\widetilde T^{-\bsigma}(g^{-1}) \bar e_m,e_n)=
\Omega_\bsigma( \bar e_m,T^\bsigma(g) e_n)\,.
\end{align}
Here $e_n(z)$ and $\bar e_m(w)=(e_m(w))^*$ are  basis vectors in the spaces
$\mathbb{V}$, $\widebar{\mathbb{V}}$~(see Eq.~(\ref{enz})). The transformations
$T^\bsigma(g),\widetilde T^{-\bsigma}(g^{-1})$ are defined by Eqs.~(\ref{Tg}),~(\ref{Ttilde}).
 }\\
\begin{proof}
Since $\Delta_k(g)\neq 0$ the functions
$T^\bsigma(g) e_n(z)$ and $\widetilde T^{-\bsigma}(g^{-1}) \bar e_m(z)$ are (anti)holomorphic
functions in the  vicinity of the point $z=1$ ($z_{ik}=0$)
\begin{align}\label{ab}
[T^{\balpha}(g)e_n](z)=\sum_{k} A_{kn} e_k(z)\,,&&
 [\widetilde T^{\balpha}(g^{-1})e_m](\bar z)=\sum_{k} \overline{e_k(z)}\,B_{mk}\,.
\end{align}
The matrices $A$ and $B$ satisfy the following relation
\begin{align}\label{A-B}
(\Omega^{-1}\, B)_{nm}=(A\,\Omega^{-1})_{nm}\,,
\end{align}
which follows immediately from the identity for the
reproducing kernel, 
$$
[T_z^\bsigma(g)\mathcal{I}^\bsigma](z,w)=
[\widetilde T^{-\bsigma}_w(g^{-1})\mathcal{I}^\bsigma](z,w),
$$ 
(see Eq.~(\ref{TtT})).
Inserting (\ref{ab}) into
(\ref{tpq}) one obtains that the latter reduces to equation (\ref{A-B}).
\end{proof}
\vskip 2mm

Further if both functions $Q(\bar z)$ and $[\widetilde T^{-\bsigma}(g^{-1})Q](\bar z)$ are
analytic at $z=1$  then one obtains
\begin{align}\label{Tad}
\Omega_\bsigma(\widetilde T^{-\bsigma}(g^{-1})Q,e_m)=\sum_n q_n \,
\Omega_\bsigma(\bar e_n,T^\bsigma(g) e_m)\,,
\end{align}
where $Q(\bar z)=\sum_n q_n \overline{e_n(z)}$.
\vskip 5mm

{\lemma \label{lemma2}
If the kernel of an operator $A$ has the form
\begin{align}\label{}
A(z,\alpha)= r(z,\bar\alpha_z)\, \mathcal{I}^\bsigma(z,\alpha)\,,
\end{align}
where the function $r(z,\bar\alpha)$ does not depend on the variables:\\[2mm]
a)\,$\alpha^*_{nj}$, $j< m$\\[2mm]
or
\\[2mm]
b) \, $(\alpha^{-1})^*_{jn}$, $j> m$\\[2mm]
then\\[2mm]
a) $A$ commutes with $z_{kj}$, $j<m$,\quad $A\,z_{kj}=z_{kj}\,A$\,,\\[2mm]
b) $A$ commutes with $z^{-1}_{jk}$, $j>m$,\quad $A\,z^{-1}_{jk}=z^{-1}_{jk}\,A$\,.
}
\vskip 2mm
\noindent
\begin{proof} 
The proofs for the cases a) and b) are similar, so we consider the case a) only.
Since the function $r(z,\bar\alpha)$ depends only on a part
of the variables $\alpha_{nj}^*$ its  expansion  in a power series
\begin{align}\label{rm}
r(z,\bar\alpha)=\sum_{ij} r_{ij}\,e_i(z)
\, \overline{e_j(\alpha)}\,
\end{align}
runs only over those basis vectors $ \bar{e}_j$ which lie in the subspaces
$\widebar{\mathbb{V}}_{h}$ with the multi-index $h=(0,\ldots,0,h_m,\ldots,h_{N-1})$.
Therefore, taking into account Eq.~(\ref{Omega}) one derives
\begin{align}\label{}
\Omega_\bsigma(r(z,\bar\alpha),P(\alpha))=\Omega_\bsigma(r(z,\bar\alpha),\Pi_m\,P(\alpha))\,,
\end{align}
where $\Pi_m$ is a projector to the subspace
$\mathbb{V}_m=\sum_{h=(0,\ldots,0,h_m,\ldots,h_{N-1})}\oplus \mathbb{V}_{h}$,
\begin{align}\label{}
[\Pi_m \, P](\alpha)=P(\alpha)\Big|_{\alpha_{kj}=0,\, j<m}%
\,.
\end{align}
Noticing that the kernel $A(z,\alpha)$ has the form $\widetilde T^{-\bsigma}(z) r(z,\bar\alpha)$
and making use of Eq.~(\ref{Tad}) one obtains
\begin{multline}\label{pr}
A  \,P(z)=
\Omega_{\bsigma}\Big(A(z,\alpha),
  P(\alpha)\Big)=\Omega_{\bsigma}\Big({\widetilde T}^{-\bsigma}(z)\,
r(z,\bar\alpha),\, P(\alpha)\Big)=\\
\sum_{ij}r_{ij}\,e_i(z)\,
\Omega_{\bsigma}\Big(\overline{e_j(\alpha)}\,, {T}^\bsigma(
z^{-1})\, P(\alpha)\Big)=\sum_{ij}r_{ij}\,e_i(z)\,
\Omega_{\bsigma}\Big(\overline{e_j(\alpha)}\,, \Pi_m\,{T}^\bsigma(
z^{-1})\, P(\alpha)\Big)\,.
\end{multline}
Taking $P(z)=z_{kj} \widetilde P(z)$, $j<m$ and noticing that
$$
\Pi_m\,{T}^\bsigma(z^{-1})\, \alpha_{kj}\widetilde P(\alpha)=z_{kj}
\Pi_m\,{T}^\bsigma(z^{-1})\, \widetilde P(\alpha)
$$
we obtain the necessary result.
\end{proof}
\vskip 2mm

The commutation relations~(\ref{Rz}) for the operators~$\mathbb{R}^{(m)}_{12}(\lambda)$ is
a simple corollary of this lemma.
The proof of the  two remaining relations~(\ref{DR}) and (\ref{p1}) is given in Appendix~\ref{AppB}.

Finally, with the help of the  relations~(\ref{Rz}) one can check that the
operator $\mathbb{R}^{(m)}(\lambda)$ %
depends only on a part
of the parameters $\bsigma$ and $\brho$ characterizing  the Verma modules, namely
\begin{align}\label{exR}
\mathbb{R}^{(m)}_{\bsigma\brho}(\lambda)=\mathbb{R}^{(m)}(\lambda,\sigma_{12},\ldots,\sigma_{m-1m},
\rho_{mm+1},\ldots,\rho_{N-1N})\,.
\end{align}
\vskip 2mm

Proving the properties~(\ref{propR}),~(\ref{Rz}),~(\ref{DR}) we have assumed that 
the representations of the $sl(N)$ algebra the factorizing operators act on are irreducible.
This condition can be relaxed. Indeed, all
relations in question give rise (and are equivalent) to certain equations for  matrix
elements of  the factorizing operators $\mathbb{R}^{(m)}$. The matrix elements
depend analytically on the parameters
specifying the representations. Thus these equations hold for
arbitrary parameters and, hence, for reducible representations.

On the basis of the obtained results we conclude that the following theorem holds:
{\theorem A $sl(N)$ invariant solution of the YBE on a tensor product of
generic highest weight representations of the $sl(N)$ algebra can be represented in the
factorized
form
\begin{align}\label{RVV-factform}
\CR_{12}(u-v)=P_{12}\,\mathbb{R}^{(1)}_{12}(u_1-v_1)
\mathbb{R}^{(2)}_{12}(u_2-v_2)
\ldots
\mathbb{R}^{(N)}_{12}(u_N-v_N)\,.
\end{align}
Here $u_i=u-\sigma_i$, $v_i=v-\rho_i$. The parameters $\bsigma,\brho$ specify
the representations $\pi^{(1)}$ and $\pi^{(2)}$, respectively. $P_{12}$ is a
permutation operator. The factorizing operators
$\mathbb{R}^{(m)}_{12}(\lambda)$ are given by Eq.~(\ref{r-kernel}) and satisfy
(at proper normalization) the relations~(\ref{propR}),~(\ref{Rz}),~(\ref{DR}).
}
\vskip 5mm

The Verma module
is irreducible if none of the
differences, $\sigma_i-\sigma_k$, $k>i$, is a positive integer. In other cases
there exists an (in)finite dimensional invariant subspace $\upsilon$,
$\upsilon\subset \mathbb{V}$. Let $\pi'$ be a restriction of $\pi^{\bsigma}$ onto the
subspace $\upsilon$, $\pi'=\pi^{\bsigma}|_{\upsilon}$, and $\pi''$ is a representation
induced on the factor space $\mathbb{V}/\upsilon$. It follows from the $RLL$ relation that
the space $\upsilon\otimes\mathbb{V}$ (we assume that the representation $\pi^{\brho}$ is irreducible)
is an invariant subspace of the operator $\mathcal{R}_{12}$. Therefore the $\mathcal{R}-$matrix
has a block-triangular form
\begin{align}\label{}
\mathcal{R}_{12}(u)=\begin{pmatrix} \mathcal{R}'_{12}(u)& \star\\
                                     0& \mathcal{R}''_{12}(u)
\end{pmatrix}\,,
\end{align}
where the diagonal blocks, $
\mathcal{R}'_{12}(u)=\mathcal{R}_{12}(u)\Big|_{\upsilon\otimes \mathbb{V}}$  and
$ \mathcal{R}''_{12}(u)=\mathcal{R}_{12}(u)\Big|_{(V/\upsilon)\otimes \mathbb{V}}$, define new
$\mathcal{R}$ matrices on the spaces $\upsilon\otimes \mathbb{V}$ and $(V/\upsilon)\otimes
\mathbb{V}$, respectively. Thus one can extract $ \mathcal{R}-$matrices for arbitrary
(non-generic) representations of the $sl(N)$ algebra by studying the $
\mathcal{R}-$matrix~(\ref{RVV-factform})  for  
reducible  generic representations.
\vskip 5mm

Let us put $u_m-v_m=\lambda$ and $u_k=v_k$, $k\neq m$ in Eq.~(\ref{RVV-factform}). These
constraints mean that the characters $\balpha$ and $\bbeta$ in the tensor product
$\pi^{\balpha}\otimes \pi^{\bbeta}$ are related. Namely, one easily finds that
$\bbeta(h)=\balpha_{m,\lambda}(h)=h_{mm}^{-\lambda}\balpha(h)$.
Taking
into account that $\mathbb{R}^{(k)}_{12}(0)=\mathbb{I}$ one derives
\begin{align}\label{RmR}
\mathcal{R}^{(m)}_{12}(\lambda)\equiv P_{12}\,\mathbb{R}^{(m)}_{12}(\lambda)=
\CR_{12}\left(\frac\lambda{N}\right)\Big|_{\bbeta=\balpha_{m,\lambda}}\,.
\end{align}
Constructing transfer matrices one refers to spaces
in the tensor product $\mathbb{V}\otimes \mathbb{V}$ as
a quantum  (first) and  auxiliary (second) spaces, respectively. Clearly, the operator
$\mathcal{R}^{(m)}_{12}(\lambda)$ is completely fixed by the spectral parameter $\lambda$
and the representation $\pi^\balpha(\equiv\pi^{\bsigma})$ on the quantum space. 
Making use of Eq.~(\ref{exR})
one gets
\begin{align}\label{}
\mathcal{R}^{(m)}_{12}(\lambda)=P_{12}\,
\mathbb{R}^{(m)}(\lambda,\sigma_{12},\ldots,\sigma_{m-1m},
\lambda+\sigma_{mm+1},\ldots,\lambda+\sigma_{N-1N})\,.
\end{align}
In the next section we study the properties of the operators defined as the trace of a
monodromy matrices constructed from the
operators $\mathcal{R}^{(m)}_{12}(\lambda)$.
%

\section{Transfer matrices and Baxter $\mathcal{Q}-$operators}\label{TmQ}

\subsection{Transfer matrices}
In this section we discuss  properties of transfer matrices for generic $sl(N)$ spin
chains. A~transfer matrix is defined as a trace of a monodromy matrix which is given by
the product of $\mathcal{R}-$operators. Since
we consider   infinite-dimensional
representations of the $sl(N)$ algebra a convergence of the trace is not guaranteed.
Except for the $N=2$ case, in order to ensure the finiteness of traces one has to
introduce  a regulator (boundary operator)~\cite{BLZ-III,BHK,Korff04,Korff06} 
which breaks $sl(N)$ symmetry down to its
diagonal subgroup. Namely, let us define a new  $\mathcal{R}-$operator by
\begin{align}\label{Rtau}
\mathcal{R}_{12}(u,\tau)\equiv \mathcal{R}_{12}(u,\tau_1,\ldots,\tau_{N-1})=
\prod_{p=1}^{N-1}\tau_p^{(H_2)_p}\, \mathcal{R}_{12}(u)\equiv \tau^{H_2} \,\mathcal{R}_{12}(u)\,.
\end{align}
The operators $H_p$ are defined in (\ref{H-p}) and
the index $2$  ( $H_2$ ) refers to the space the operator acts on.
The operators
$\mathcal{R}_{12}(u,\tau)$ obey the YBE
\begin{align}\label{YBp}
\mathcal{R}_{12}(u,\tau)\,\mathcal{R}_{13}(v,\tau)\,\mathcal{R}_{23}(v-u)=
\mathcal{R}_{23}(v-u)\,\mathcal{R}_{13}(v,\tau)\,\mathcal{R}_{12}(u,\tau)\,.
\end{align}
We define a transfer matrix by~\footnote{In Refs.~\cite{Korff04}
the transfer matrix with an insertion of a boundary operator was defined as follows
$\widetilde T(u,\tau)=\tr_{\rho} (\tau^{H_0} R_{10}(u)\ldots R_{L0}(u))$. It is easy to
check that these two definitions are essentially the same
$T(u,\tau)=\tau^{A}\widetilde T(u,\tau^L)\tau^{-A}$, where $A=\sum_{k=1}^{L-1}(L-k) H_{k}$.
}
\begin{align}\label{}
    \label{deft}
{\sf T}_{\brho}(u,\tau) =\tr_{\brho} \Big\{\CR_{10}(u,\tau)\ldots \CR_{L0}(u,\tau)\Big\}\,.
\end{align}
The index $\brho$ specifies a representation  of $sl(N)$
algebra, $\pi^{\brho}(=\pi^{\bbeta})$, on the auxiliary space. 
We will also label a transfer matrix by a highest weight of the  representation
in the auxiliary space $\blambda=(\lambda_1,\ldots,\lambda_{N-1})$,
$\lambda_k=1-\rho_k+\rho_{k+1}$, i.e.
\begin{align}\label{tlb}	
{\sf T}_{\blambda}(u,\tau)\equiv {\sf T}_{\bbeta}(u,\tau)\equiv {\sf T}_{\brho}(u,\tau)\,.
\end{align}

Let us note that for $\tau<1$ the factor $\tau^{H_2}$ improves a
convergence of the trace since the eigenvalues of the operators $H_p$,
$p=1,\ldots,N-1$, (see Eq.~(\ref{H-p})) are positive integers.
Thus, provided that the trace exists, Eq.~(\ref{deft}) defines an
operator on the tensor product of Verma modules
$\mathbb{V}_1\otimes\ldots\otimes \mathbb{V}_{L}$. The transfer matrices form a commutative family of
operators
\begin{align}\label{}
[{\sf T}_{\brho_1}(u,\tau),{\sf T}_{\brho_2}(v,\tau)]=0\,.
\end{align}
It is easy to see that  the transfer matrix ${\sf
T}_{\brho}(u,\tau)$ commutes
 also with the total Cartan generators,
namely $[{\sf T}_{\brho}(u,\tau), H_{p}]=0$, $p=1,\ldots,N-1$,
where $H=H^{(1)}+\ldots+H^{(L)}$.

We will
consider  homogeneous spin chains only, i.e. assume that the
representations of $sl(N)$ algebra on the quantum space at each
site are equivalent,
$\pi^{\bsigma_1}=\pi^{\bsigma_2}=\ldots=\pi^{\bsigma_L}\equiv
\pi^{\bsigma}$.

\subsection{Baxter $\mathcal{Q}-$operators}
Let us define  special transfer matrices, the Baxter $Q$-operators, constructed from  the operators
$\mathcal{R}^{(m)}$,~(\ref{RmR}). Namely, similarly to~(\ref{Rtau}) we put
\begin{align}\label{}
\mathcal{R}^{(m)}_{12}(u,\tau)=\tau^{H_2}\, \mathcal{R}^{(m)}_{12}(u)\,
\end{align}
and
\begin{align}\label{defQ}
\mathcal{Q}_k(u+\sigma_k,\tau)=
\tr_{0} \Big\{\CR_{10}^{(k)}(u,\tau)\ldots \CR_{L0}^{(k)}(u,\tau)\Big\}\,.
\end{align}
Taking into account Eq.~(\ref{RmR}) one finds that $\mathcal{Q}_k(u)$ is given by
the transfer matrix for a special choice of the auxiliary space
\begin{align}\label{Q=T}
\mathcal{Q}_k(u+\sigma_k,\tau)={\sf T}_{\balpha_{k,u}}(u/N,\tau)={\sf
T}_{\blambda-u(e_k-e_{k-1})}(u/N,\tau)\,,
\end{align}
where the vector  $e_{k}$ is defined by $(e_k)_n=\delta_{kn}$, and $\blambda$ is the
highest weight of the
representation in the quantum space, $\lambda_k=\sigma_{k+1}-\sigma_k+1$.

The Baxter  operators $\mathcal{Q}_k(u,\tau)$ act on the quantum space of a model
and commute with each other
\begin{align}\label{}
[\mathcal{Q}_k(u,\tau),\mathcal{Q}_m(v,\tau)]=0\,.
\end{align}
One easily finds that the Baxter operators satisfy the following normalization condition
\begin{align}\label{QPH}
\mathcal{Q}_k(\sigma_k,\tau)=\mathcal{P}\,\tau^{H}\,,
\end{align}
where $H=H^{(1)}+\ldots+H^{(L)}$ and $\mathcal{P}$ is the operator of cyclic permutation,
\begin{align}
\mathcal{P}\, f(z_1,\ldots,z_L)=f(z_L,z_1,\ldots,z_{L-1}).
\end{align}
Let us note also that $[\mathcal{Q}_k(u),\mathcal{P}]=[\mathcal{Q}_k(u),H]=0$.
\vskip 5mm

Now we are going to prove that Eq.~(\ref{defQ}) gives rise to a well defined operator on
the quantum space. To this end it is necessary to show that the trace over an infinite
dimensional auxiliary space converges.
Let
$\mathbb{Q}_k(u,\tau)$ be a monodromy matrix,
\begin{align}\label{}
\mathbb{Q}_k(u,\tau)=
\mathcal{R}_{10}^{(k)}(u,\tau)\ldots\mathcal{R}_{L0}^{(k)}(u,\tau)\,,
\end{align}
and $[\mathbb{Q}_k(u,\tau)]_{m_1\ldots m_L,n}^{m'_1\ldots m'_L,n'}$  its matrix
elements in the basis $E_{\vec{m},n}$,
\begin{align*}
E_{\vec{m},n}=&e_n(w)\prod_{k=1}^L e_{m_k}(z^{(k)})\,,\\
\mathcal{Q}_k(u\tau)E_{\vec{m},n}=&
[\mathbb{Q}_k(u,\tau)]_{m_1\ldots m_L,n}^{m'_1\ldots m'_L,n'}
E_{\vec{m'},n'}\,.
\end{align*}
 Here we keep the variable $w$ for the auxiliary space and $z^{(k)}$
for the quantum space in  $k-$th site. The basis vectors $e_n(z)$
 are defined by Eq.~(\ref{enz}).

The trace of the monodromy matrix $\mathbb{Q}_k(u,\tau)$  written explicitely takes the form
\begin{align}\label{sumQ}
\sum_{\boldsymbol{n}=(n_1,n_2,\ldots,n_L)} [\mathcal{R}_{10}^{(k)}]_{m_1,n_1}^{m'_1 n_L}
[\mathcal{R}_{20}^{(k)}]_{m_2,n_2}^{m'_2 n_1}\ldots
[\mathcal{R}_{L0}^{(k)}]_{m_L,n_{L}}^{m'_L n_{L-1}}\,.
\end{align}
We recall that each summation index, $n_k$ is multi-index, $n_k=\{(n_k)_{ij}, \,\, i>j\}$.
Let us introduce notations, $\boldsymbol{n}=(n_1,n_2,\ldots,n_L)$, and
 $\boldsymbol{n}_{ij}=((n_1)_{ij},(n_2)_{ij},\ldots,(n_L)_{ij})$.
First of all, we show that
all summation  indices $\boldsymbol{n}_{ij}$ except
$\boldsymbol{n}_{k+1k},\ldots,\boldsymbol{n}_{Nk}$
 vary in a finite range, while the indices $\boldsymbol{n}_{k+1k},\ldots,\boldsymbol{n}_{Nk}$
can be arbitrarily large.
To this end we examine an action of the operator $\mathcal{R}_{l0}^{(k)}$ on the basis vector
$e_{m_l}(z_l)\otimes e_{n_l}(w)$. For brevity we skip the index $l$, i.e.
$z_l\to z$, $m_l\to m$, etc. Then taking into account Eqs.~(\ref{Rz}) one finds
\begin{multline}\label{}
\mathcal{R}_{l0}^{(k)} e_{m}(z)\otimes e_{n}(w)=P_{zw}
\left(\prod_{j<k,i}w_{ij}^{n_{ij}}\right)\,
\mathbb{R}_{l0}^{(k)}\,
\left(e_{m}(z)\otimes \prod_{i>j\geq k}w_{ij}^{n_{ij}}\right)=\\
\left(\prod_{j<k,i}z_{ij}^{n_{ij}}\right)\,
\mathcal{R}_{l0}^{(k)}\,
\left(e_{m}(z)\otimes \prod_{i>j\geq k}w_{ij}^{n_{ij}}\right)\,.
\end{multline}
From here one concludes that the indices $n_{ij}$ for $j<k$ are restricted from above by $m'_{ij}$,
$n_{ij}\leq m'_{ij}$, i.e. $\boldsymbol{n}_{ij}\leq \boldsymbol{m}'_{ij}$ for $j<k$.

To prove that $\boldsymbol{n}_{ij}$ is restricted for $j>k$ we use the relation~(\ref{DR}).
It takes the form
\begin{align}
D^{w}_{p+1,p}\mathcal{R}_{l0}^{(k)}=\mathcal{R}_{l0}^{(k)}D^{z}_{p+1,p}, \qquad p>k.
\end{align}
Since for a given $m$ the operators $(D^z_{p+1,p})^{M_p}$, where $M_p$ is some number, nullify
the vector $e_m(z)$ one derives that
$(D^w_{p+1,p})^{M_p} E_{mn}(z,w)=0$ for $p>k$,
 where  $E_{mn}(z,w)=\mathcal{R}_{l0}^{(k)}(e_m(z)\otimes e_n(w))$.
The function $E_{mn}$ satisfying these conditions is a polynomial of finite degree (which
depends on $M_p$)
 in $w_{ip}$, $p>k$ (see Ref.~\cite{Zhelobenko}, chapter X) hence,
$\boldsymbol{n}_{ij}\leq \boldsymbol{M'}_{ij}$, for $j>k$.

Thus we have shown that the summation over $n_p=\{(n_p)_{ij}\}$, $p=1,\ldots,L$
in~(\ref{sumQ}) goes in a finite
range for all $(n_p)_{ij}$ except $(n_p)_{k+1,k},\ldots, (n_p)_{N,k}$. Let us also notice
that all summation indices, $(n_p)_{ik}$ are of the same order, the difference
$(n_p)_{ik}-(n_L)_{ik}=(q_p)_{ik}$ being finite when $(n_L)_{ik}$ goes
to infinity. It will be shown in Appendix~\ref{R-asymptotic} that for the matrix element
$[\mathcal{R}_{l0}^{(k)}(u)]_{mn}^{m'n'}$ in the limit $n_{ik}\to \infty$,
$i=k+1,\ldots, N$,  all other variables being fixed, the following estimate holds
\begin{align}\label{estR}
|[\mathcal{R}_{l0}^{(k)}(u)]_{mn}^{m'n'}|< C(u)
h_{k+1}^{a_{k+1}}
\ldots h_{N}^{a_{N}}\,,
\end{align}
where $h_{k}=\sum_{i=k+1}^{N} n_{ik}+1$ and $C(u)$, $a_k,\ldots, a_N$ are some constants, which
depends on $\bsigma, m_{ij},\ldots$. Since $\tau^{H_2}\sim \tau_k^{h_k}\ldots
\tau_{N-1}^{h_{N-1}}$ the estimate~(\ref{estR}) ensures that the series
in~(\ref{sumQ}) converges absolutely $\boldsymbol{\tau} < 1$. 
Thus we have proven that Eq.~(\ref{defQ}) provides
a definition of an operator on the tensor product of Verma modules.

Let us note that the trace for the operator $\mathcal{Q}_N$ is given by  a finite
sum. Therefore,  the Baxter operator $\mathcal{Q}_N(u,\tau)$ has a
finite limit at $\tau\to 1$. The operators $\mathcal{Q}_k(u,\tau)$, $k<N$, could be
singular in this limit.  However, as it follows from the above discussion, the operator
$\mathcal{Q}_k(u,\tau)$ has a finite limit at  $\tau_i\to 1$,
$i\neq k$ and $\tau_k<1$ is fixed.

\begin{figure}[t]
\psfrag{RR}[cc][cc][1.2]{$A^{n'm'}_{nm}=$}
\psfrag{i}[cc][cc]{$n'$}
\psfrag{ia}[cc][cc]{$n$}
\psfrag{0}[cc][cc]{$m'$}
\psfrag{0a}[cc][cc]{$m$}
\psfrag{R}[cc][cc]{$A$}
\centerline{\includegraphics[width=7cm]{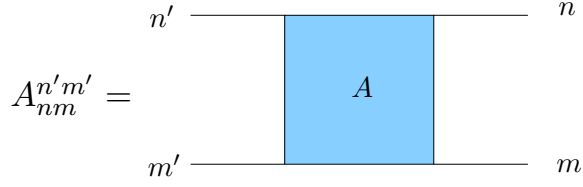}}\hskip 0.5cm
\caption{The graphical representation for the matrix element of an operator.}
\label{box}
\end{figure}
\begin{figure}[t]
\psfrag{A}[cc][cc]{$A$}
\psfrag{B}[cc][cc]{$B$}
\psfrag{0}[cc][cc]{$m'''$}
\psfrag{i}[cc][cc]{$n''$}
\psfrag{c}[cc][cc]{$n$}
\psfrag{d}[cc][cc]{$m$}
\centerline{\includegraphics[width=7cm]{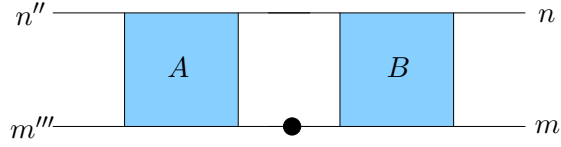}}\hskip 0.5cm
\caption{The graphical representation for the sum $\sum_{n',m',m''}A_{n'm''}^{n''m'''}
[\tau^{H}]_{m'}^{m''}\, B_{n m}^{n'm'}$.}
\label{example}
\end{figure}

\subsection{Factorized form of  transfer matrix}

The transfer matrix~(\ref{deft}) for a chosen quantum space depends on $N$ complex
parameters: the spectral parameter $u$ and $N-1$ parameters, $\rho_{k}-\rho_{k+1}$,
$k=1,\ldots,N-1$, specifying the
representation on auxiliary space. The Baxter  $\mathcal{Q}-$operators depend only on a
spectral parameter. The number of independent parameters in the transfer matrix matches the
number of  spectral parameters in a product of $N$ Baxter operators. Below we show that the
following statement holds

{\theorem \label{Factorization} The transfer matrix ${\sf T}_{\brho}(u,\tau)$, $\tau
<1$, is factorized into the product of the Baxter $\mathcal{Q}-$operators
\begin{align}\label{TQ}
\mathsf{T}_\brho(u,\tau)=\mathcal{Q}_1(u+\rho_1,\tau)\,
\left(\mathcal{P}\tau^{{H}}\right)^{-1}\,
\mathcal{Q}_2(u+\rho_2,\tau)\, 
\ldots
  \left(\mathcal{P}\tau^{{H}}\right)^{-1}\,\,
\mathcal{Q}_N(u+\rho_N,\tau)\,,
\end{align}
where $H=H^{(1)}+\ldots+H^{(L)}$.
}
\begin{figure}[t]
\psfrag{a3}[cc][cc]{$A_3$}
\psfrag{b3}[cc][cc]{$B_3$}
\psfrag{a1}[cc][cc]{$A_1$}
\psfrag{b1}[cc][cc]{$B_1$}
\psfrag{a2}[cc][cc]{$A_2$}
\psfrag{b2}[cc][cc]{$B_2$}
\psfrag{1}[cc][cc]{$m'_1$}
\psfrag{2}[cc][cc]{$m'_2$}
\psfrag{N}[cc][cc]{$m'_3$}
\psfrag{Na}[cc][cc]{$m_3$}
\psfrag{1a}[cc][cc]{$m_1$}
\psfrag{N1a}[cc][cc]{$m_2$}
\centerline{\includegraphics[width=12cm]{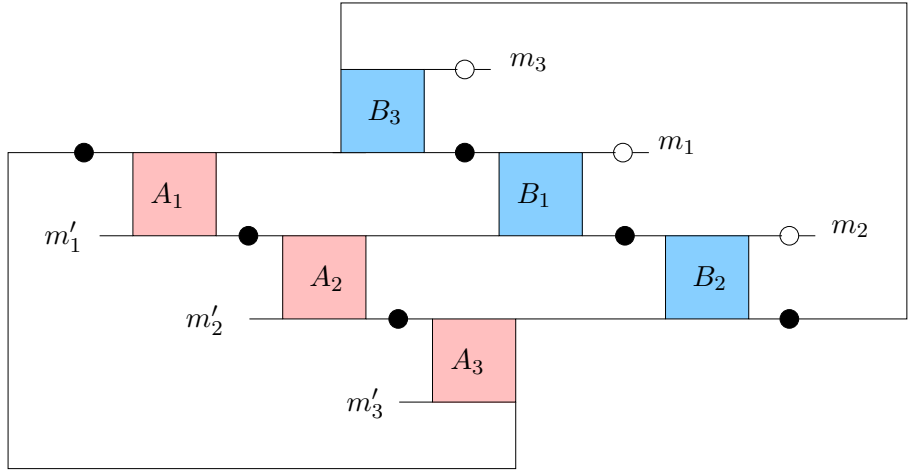}}\hskip 0.5cm
\caption{The graphical representation for the reduced transfer matrix ${\sf T}_k$,~the
r.h.s of Eq.~(\ref{Tk}).}
\label{QQ}
\end{figure}

\begin{figure}[t]
\psfrag{a}[cc][cc]{$m_1$}
\psfrag{b}[cc][cc]{$m_2$}
\psfrag{c}[cc][cc]{$m_3$}
\psfrag{b1}[cc][cc]{$B_2$}
\psfrag{a2}[cc][cc]{$A_3$}
\psfrag{b2}[cc][cc]{$B_3$}
\psfrag{i}[cc][cc]{$B_j$}
\psfrag{k}[cc][cc]{$A_{j+1}$}
\psfrag{3}[cc][cc]{$m'_3$}
\psfrag{1}[cc][cc]{$m'_1$}
\psfrag{2}[cc][cc]{$m'_2$}
\psfrag{=}[cc][cc][1.2]{$=$}
\centerline{\includegraphics[width=14cm]{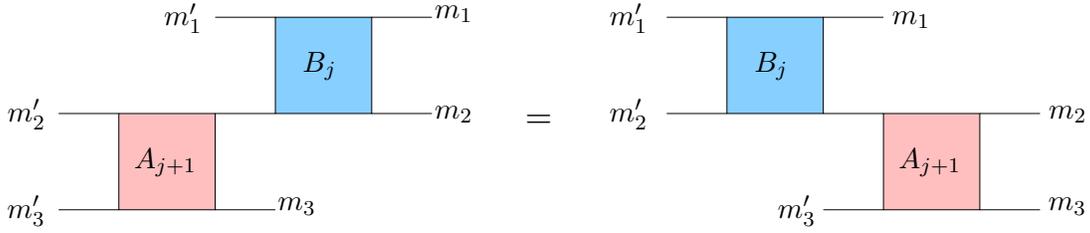}}\hskip 0.5cm
\caption{The graphical representation of the permutation identity~(\ref{R-R}).}
\label{RR-diagram}
\end{figure}

\noindent
\begin{proof} 
The proof of (\ref{TQ}) relies on the commutation relations~(\ref{R-R}).
Let ${\sf T}_k(u,\tau)$ will be the transfer matrix for a special choice of the
auxiliary space
$$
\bbeta_k=\prod_{j=1}^k \Delta_j^{1-\rho_{jj+1}}\,
\Delta_k^{1-\rho_k+\sigma_k-u}
\,\prod_{j=k+1}^{N-1} \Delta_j^{1-\sigma_{jj+1}}\,,
$$
where $\boldsymbol{\sigma}=(\sigma_1,\ldots,\sigma_N)$  are
the parameters specifying the representation
 on the quantum space. 
The transfer matrix ${\sf T}_k(u,\tau)$ can be represented in the form
\begin{align}\label{Tk}
{\sf T}_k(u,\tau)=\tr_{\bbeta_k}\mathcal{R}^{(1k)}_{10}(u,\tau)
\ldots \mathcal{R}^{(1k)}_{L0}(u,\tau)\,.
\end{align}
The reduced operator $\mathcal{R}^{(1k)}$ is defined as follows
\begin{align}
\mathcal{R}^{(1k)}_{j0}(u-v,\tau)=\tau^{H_0}P_{j0}
\mathbb{R}^{(1)}_{j0}(u_1-v_1)\ldots\mathbb{R}^{(k)}_{j0}(u_k-v_k)
\equiv\tau^{H_0}\,P_{j0}\,\mathbb{R}^{(1k)}_{j0}(u-v,\tau)\,,
\end{align}
where $u_j=u-\sigma_j$, $v_j=v-\rho_j$. Let us show that
\begin{align}\label{tkk}
{\sf T}_k(u,\tau)={\sf T}_{k-1}(u,\tau)\,\left(\mathcal{P}\tau^{{H}}\right)^{-1}\,
\mathcal{Q}_k(u+\rho_k,\tau)\,.
\end{align}
Obviously, the factorization formula~(\ref{TQ}) is a simple corollary of  this result.
To prove Eq.~(\ref{tkk}) let us put $A_j=\mathbb{R}^{(1k-1)}_{j0}(u-v)$ and
$B_j=\mathbb{R}^{(k)}_{j0}(u_k-v_k)$, i.e.
\begin{align}\label{}
\mathcal{R}^{(1k)}_{j0}(u-v,\tau)=\tau^{H_0} P_{j0}\,A_j\,B_j.
\end{align}
Matrix elements of the operators on both sides in Eq.~(\ref{tkk}) are given by some
sums and one has to show that they are equal. To this end it is convenient to use a
graphical representation for the sums. Let us denote the matrix element $A^{n'm'}_{nm}$
of the operator
$A_j(B_j)$ by a box with four legs as shown in Fig.~\ref{box}.
The line connecting two boxes will imply a summation over the corresponding index. The
operators $\tau^{H}$ ( $\tau^{-H}$ ) will be denoted by an insertion of black (white)
circle in the corresponding line. As an example, we have given the diagrammatic 
representation for the sum
$\sum_{n',m',m''}A_{n'm''}^{n''m'''}
[\tau^{H}]_{m'}^{m''}\, B_{n m}^{n'm'}$  in
Figure~\ref{example}.
\begin{figure}[t]
\psfrag{a}[cc][cc]{$A_1$}
\psfrag{b}[cc][cc]{$B_1$}
\psfrag{a1}[cc][cc]{$A_2$}
\psfrag{b1}[cc][cc]{$B_2$}
\psfrag{a2}[cc][cc]{$A_3$}
\psfrag{b2}[cc][cc]{$B_3$}
\psfrag{1}[cc][cc]{$m'_1$}
\psfrag{2}[cc][cc]{$m'_2$}
\psfrag{3}[cc][cc]{$m'_3$}
\psfrag{1a}[cc][cc]{$m_1$}
\psfrag{2a}[cc][cc]{$m_2$}
\psfrag{3a}[cc][cc]{$m_3$}
\centerline{\includegraphics[width=14cm]{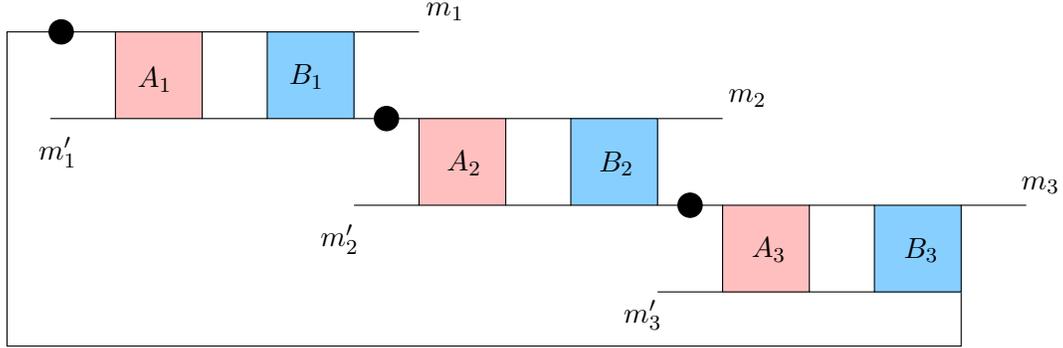}}\hskip 0.5cm
\caption{The graphical representation for the reduced transfer matrix ${\sf T}_k$, the
l.h.s. of Eq.~(\ref{Tk}).}
\label{T-diagram}
\end{figure}

The graphical representations for the r.h.s. and l.h.s. of Eq.~(\ref{tkk})
are shown in Figs.~\ref{QQ} and~\ref{T-diagram}, respectively.
To obtain the diagram shown
in Fig.~\ref{QQ} we have used
 the commutativity of the Baxter operators with the total diagonal generators,
$H=H^{(1)}+\ldots+H^{(L)}$ and represented the r.h.s. of  Eq.~(\ref{tkk}) as
$
T_{k-1}(u,\tau)\,\mathcal{P}^{-1}\,
\mathcal{Q}_k(u+\rho_k,\tau)\, \tau^{-{H}}
$.

We have already shown that the trace in Eq.~(\ref{defQ}) 
converges absolutely for $\tau<1$. Let as assume now that the factorization
formula~(\ref{tkk}) holds for the transfer matrices ${\sf T}_{k}(u,\brho)$,
$k=2,\ldots,p-1$ and that the traces for  ${\sf T}_{k}(u,\brho)$, $k\leq p-1$ converge absolutely.
Then it can be shown that Eq.~(\ref{tkk}) holds for  $k=p$
and the corresponding trace converges absolutely.
First of all,   let  notice that the summation over indices
$k_1,\ldots k_L$ in
$$
\sum_{k_1,\ldots,k_L}[{\sf T}_{k-1}]^{m'_1,\ldots,m'_L}_{k_1,\ldots,k_L}
[\mathcal{P}^{-1}Q_{k}\tau^{-H}]_{m_1,\ldots,m_L}^{k_1,\ldots,k_L}
$$
goes in a finite range. The blocks $A_{j}$ and $B_{j+1}$ can be interchanged with the
help of the commutation relation~(\ref{R-R}) whose graphical
form shown in Fig.~\ref{RR-diagram}.
Using this identity and taking into account that $ \tau^{H_j+H_0}B_j\tau^{-H_j}=B_j\tau^{H_0}$
\begin{figure}[h]
\psfrag{3}[cc][cc]{$m_3$}
\psfrag{1}[cc][cc]{$m_1$}
\psfrag{R}[cc][cc]{$B_j$}
\psfrag{=}[cc][cc][1.2]{$=$}
\centerline{\includegraphics[width=8cm]{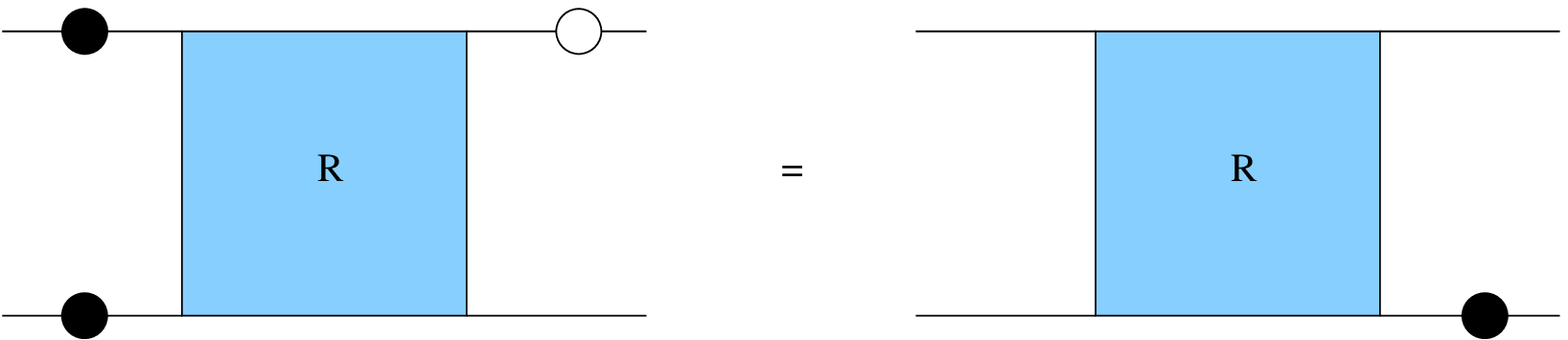}}\hskip 0.5cm
\label{Btau}
\end{figure}

\noindent
one can transform the sum depicted by the diagram in Fig.~\ref{QQ} into the sum
in Fig.~\ref{T-diagram}.

It is clear that the trace for ${\sf T}_k$ will converge
absolutely, if it is the case for ${\sf T}_{k-1}$ and~$\mathcal{Q}_k$. Since ${\sf
T}_1(u,\tau)=\mathcal{Q}_1(u+\rho_1,\tau)$ the trace for ${\sf T}_1$ converges
absolutely. Therefore  the factorization formula (\ref{tkk}) and the absolute convergence
of the traces for ${\sf T}_k$ for $k>1$ will follow by induction over $k$. 
This completes the proof of the theorem.
\end{proof}

\subsection{Fusion relations}
It was long known that the transfer matrices satisfy a set of functional relations which
are usually referred to as fusion relations. The fusion relations for the compact spin
chains were thoroughly studied in the literature, see
e.g. Refs.~\cite{KRS,KR82,Kuniba,Tsuboi97,Reshetikhin83,BR89,Zabrodin}.
Below we show how the
representation~(\ref{TQ}) can be used to obtain
some functional relations for the transfer matrices.

Let us consider the product of two transfer matrices, ${\sf
T}_\brho(u,\tau)$ and ${\sf T}_\bomega(v,\tau)$.
Both of them can be represented in
the factorized form~(\ref{TQ}). Since all operators in~(\ref{TQ}) commute with each  other
one can interchanges the Baxter operators $\mathcal{Q}_k(u+\rho_k)$ and $\mathcal{Q}_k(v+\omega_k)$.
The new products can be identified as the transfer matrices, e.g.
\begin{align}\label{}
\left(\mathcal{P\tau^{H}}\right)^{-(N-1)}\mathcal{Q}_k(v+\omega_k)\prod_{j\neq
k}\mathcal{Q}_j(u+\rho_j)
={\sf T}_{\brho'}(u',\tau)\,,
\end{align}
where $u'=u-\delta/N$, $\rho'_j=\rho_j+\delta/N$, for $j\neq k$ and
$\rho'_k=\rho_k-\delta\left(1-1/N\right)$, $\delta=u-v+\rho_k-\omega_k$.
Therefore one obtains the following relations
\begin{align}\label{TT}
{\sf T}_\brho(u,\tau)\,{\sf T}_\omega(v,\tau)={\sf T}_{\brho'}\left(u-\delta/N,\tau\right)\,
{\sf T}_{\omega'}\left(v+\delta/N,\tau\right)\,,
\end{align}
where $\omega'_j=\omega_j-\delta/N$, $j\neq q$ and
$\omega'_k=\omega_k+\delta\left(1-1/N\right)$.
Changing the notation
${\sf T}_\brho(u,\tau)\to {\sf T}_\blambda(u,\tau)={\sf T}_{\lambda_1,\ldots,\lambda_{N-1}}(u,\tau)$
(${\sf T}_\bomega(v,\tau)\to {\sf T}_\bmu(v,\tau)$)
where $\blambda$ ($\bmu$) is the highest weight in the auxiliary space,
$\lambda_k=\rho_{k+1}-\rho_k+1$,
 one rewrites  relation~(\ref{TT}) in the form
\begin{align}\label{TT-1}
{\sf T}_\blambda(u,\tau)\,{\sf T}_\bmu(v,\tau)=
{\sf T}_{\blambda-\delta_k(\be_{k-1}-\be_k)}(u-\delta_k/N,\tau)\,{\sf
T}_{\bmu+\delta_k(\be_{k-1}-\be_k)}(v+\delta_k/N,\tau)\,,
\end{align}
where $\be_k$ is $N-1$ dimensional vector $(\be_k)_i=\delta_{ik}$ and
\begin{align}\label{}
\delta_k=u-v+\sum_{p=k}^{N-1}(\mu_p-\lambda_p)+\frac1N\sum_{p=1}^{N-1} p(\mu_p-\lambda_p)\,.
\end{align}
The fusion relations~(\ref{TT}),~(\ref{TT-1}) remain valid and for inhomogeneous spin
chains.


\subsection{Inhomogeneous spin chains}
Let us  explore
modifications  which appear in a general case of inhomogeneous spin chains with
impurities (for the $sl(2)$ case see Ref.~\cite{SD-2}). The transfer matrix is defined as
\begin{align}\label{Txi}
{\sf T}_\brho(u,\tau)=\tr_{\brho}
\mathcal{R}_{10}(u+\xi_1,\tau)\mathcal{R}_{20}(u+\xi_2,\tau)\ldots \mathcal{R}_{L0}(u+\xi_L,\tau)\,,
\end{align}
where
$\{\xi_1,\ldots,\xi_L\}$
are  impurity parameters and the quantum space in  $k-th$ site carries
the representation $\pi^{(k)}$ of the $sl(N)$ algebra.

As in the case of the homogeneous spin chain  the transfer matrix~(\ref{Txi}) can be
represented in the form~(\ref{TQ}). However,  properties of the factorizing
$\mathcal{Q}-$operators
change drastically.
First of all, let us notice that in the case of an inhomogeneous chain it is not possible to
choose  an auxiliary space representation, $\pi^{\brho}$, such that the operators
$\mathcal{R}^{k}_{n0}(u)$ map $\pi^{\brho}\otimes \pi^{\bsigma_{n}}$ onto itself,
simultaneously for all $n$. 
Therefore, the quantum numbers of the auxiliary spaces at different sites
are different.

Let us consider the monodromy matrix constructed from the 
operators $\mathcal{R}^{(k)}_{\bsigma_j\brho_j}(u,\tau)$
\begin{align}\label{Tzeta}
\mathbb{Q}_k(u,\tau)=
\mathcal{R}^{(k)}_{\bsigma_1\brho_1}(u+\zeta_1,\tau)
\ldots\mathcal{R}^{(k)}_{\bsigma_L\brho_L}(u+\zeta_L,\tau)\,,
\end{align}
where $\bsigma_j$ and $\brho_j$ are
the quantum numbers of the quantum and auxiliary spaces at the $j-$th site and
 $\zeta_j$ are the impurities. In  general, the monodromy matrix
intertwines the representations with different quantum numbers
\begin{align}\label{Qpi}
\mathbb{Q}_k(u,\tau)\,\pi^{\brho}\otimes\pi^{\bsigma_1}\otimes\ldots\otimes\pi^{\bsigma_L}=
 \pi^{\tilde\brho}\otimes\pi^{\tilde\bsigma_1}\otimes\ldots\otimes\pi^{\tilde\bsigma_L}\,
\mathbb{Q}_k(u,\tau),
\end{align}	
 where we put $\brho\equiv\brho_L$. To be precise, for $\tau\neq 1$ the above relation
holds only for the generators from the Cartan subalgebra.
Provided that the representations $\pi^{\brho}$ and $\pi^{\tilde\brho}$ coincide
(${\brho}={\tilde\brho}$), the trace of the monodromy matrix ${\sf
Q}_k(u,\tau)=
\tr\mathbb{Q}_k(u,\tau) $
  intertwines the Cartan generators of the representations
$\pi^{\bsigma_1}\otimes\ldots\otimes\pi^{\bsigma_L}$ and 
$\pi^{\tilde\bsigma_1}\otimes\ldots\otimes\pi^{\tilde\bsigma_L}$.
The condition  ${\brho}={\tilde\brho}$ fixes the quantum numbers, $\brho_n$, of all
auxiliary spaces in~(\ref{Tzeta}). We recall that
$$
\mathbb{R}^{k}_{10}(u):\pi^{\balpha}\otimes \pi^{\bbeta}\to
\pi^{\balpha_{k,u}}\otimes \pi^{\bbeta_{k,-u}}\,,
$$
with $\balpha_{k,u}(h)=h_{kk}^{-u}\,\balpha(h)$,
$\bbeta_{k,-u}(h)=h_{kk}^{u}\,\bbeta(h)$ (correspondingly,
$\mathcal{R}^{(k)}_{10}(u)=P_{12}\mathbb{R}^{k}_{10}(u)$ maps
$\pi^{\balpha}\otimes \pi^{\bbeta}\to\pi^{\bbeta_{k,-u}}\otimes\pi^{\balpha_{k,u}}$).
Thus  the parameter $\brho_n$ in Eq.~(\ref{Tzeta}) 
is determined  by the characters $(\balpha_{n-1})_{k,u+\zeta_{n-1}}$.
In its turn the quantum numbers $\tilde\bsigma_n$,  Eq.~(\ref{Qpi}), correspond
 to the character 
$$
\tilde\balpha_n(h)=h_{kk}^{\zeta_n-\zeta_{n+1}}\,\balpha_{n+1}(h).
$$
Let us notice  that the character
$\tilde\balpha_n(h)$ does not depend on the spectral parameter $u$ and is determined by the
impurity parameters $\zeta$ only.
Since one can always assume that $\sum
\zeta_k=0$  the operator ${\sf Q}_k(u,\tau)$  
is uniquely determined by two sets of parameters, $\Sigma=(\bsigma_1,\ldots,\bsigma_L)$ and
$\widetilde\Sigma=(\tilde\bsigma_1,\ldots,\tilde\bsigma_L)$,
\begin{align}\label{}
{\sf Q}_k(u,\tau|\zeta,\Sigma)={\sf Q}_k(u,\tau|\widetilde\Sigma,\Sigma)\,.
\end{align}
We will not display spins $\Sigma,\widetilde\Sigma$ assuming always that the operators are
multiplied in a covariant way
\begin{align}\label{Qsigma}
{\sf Q}_k(u,\tau) {\sf Q}_j(v,\tau)=
{\sf Q}_k(u,\tau|\widetilde{\widetilde\Sigma},\widetilde\Sigma)
{\sf Q}_j(v,\tau|\widetilde\Sigma,\Sigma)\,.
\end{align}

Having put  the impurities parameters in Eq.~(\ref{Tzeta})
to $\zeta_j=\xi_j-(\sigma_j)_k$  we, finally, define the operator ${\sf Q}_k$ as follows  
\begin{align}\label{sfQ}
{\sf Q}_k(u,\tau)=\tr
\mathcal{R}^{(k)}_{\bsigma_1\brho_1}(u+\xi_1-(\sigma_1)_k,\tau)\ldots
\mathcal{R}^{(k)}_{\bsigma_L\brho_L}(u+\xi_L-(\sigma_L)_k,\tau)
\,,
\end{align}
where the representation $\pi^{\brho_n}\equiv \pi^{\bbeta_n}$ on the auxiliary space at $n-$th
site corresponds to the character $\bbeta_n=
(\balpha_{n-1})_{k,u+\xi_{n-1}-(\sigma_{n-1})_k}$.
Quite similar to a homogeneous spin chain one can show that the  transfer matrix for an
inhomogeneous spin chain can be
represented in the factorized form
\begin{align}\label{TQi}
{\sf T}_{\brho}(u,\tau)=\tau^{-(N-1)H}
{\sf Q}_1(u+\rho_1,\tau)\mathcal{P}^{-1}{\sf Q}_2(u+\rho_2,\tau)
\mathcal{P}^{-1}\ldots\mathcal{P}^{-1}{\sf Q}_N(u+\rho_N,\tau)\,.
\end{align}
The operators ${\sf Q}_k(u,\tau)$ satisfy the following 
relations~\footnote{It should be noted that, despite appearance, Eq.~(\ref{QcQ}) is
not a commutation relation. Indeed, if written explicitly Eq.~(\ref{QcQ})
becomes
${\sf Q}_k(u,\tau|\widetilde{\widetilde\Sigma},\widetilde\Sigma)
{\sf Q}_k(v,\tau|\widetilde\Sigma,\Sigma)=
{\sf Q}_k(v,\tau|\widetilde{\widetilde{\Sigma}},\Sigma')
{\sf Q}_k(u,\tau|\Sigma',\Sigma)$.
} 
\begin{align}\label{QcQ}
{\sf Q}_k(u,\tau)\,{\sf Q}_k(v,\tau)=&{\sf Q}_k(v,\tau)\,{\sf Q}_k(u,\tau)\,,\\
\label{QcQ2}
{\sf Q}_k(u,\tau)\,\mathcal{P}^{-1}\,{\sf Q}_n(v,\tau)=&{\sf Q}_n(v,\tau)\,\mathcal{P}^{-1}\,
{\sf Q}_k(u,\tau)\,,\qquad\text{for } k\neq n\,,
\end{align}
which follow from the properties of the factorizing operators, Eqs.~(\ref{propR}).
These relations allow to rearrange the ${\sf Q}_k$ operators in the product of ${\sf T}$
matrices in an arbitrary order. 
However, the operator  ${\sf Q}_k$ alone cannot be considered as   a ``good'' operator on
the quantum space of the model.
Only the product of all operators ${\sf Q}_k$ has the necessary invariance properties.
Therefore,
it is reasonable to identify the Baxter $\mathcal{Q}_k-$operator with the transfer
matrix~(\ref{TQi}), where only one
operator ${\sf Q}_k$ depends on a spectral parameter.
Namely, let  $\bw=(w_1,\ldots,w_N)$
and $\bmu=(\mu_1,\ldots,\mu_{N-1})$, where $\mu_k=w_{k+1}-w_k+1$.
We define the Baxter
operator as follows
\begin{align}\label{Qw}
\mathcal{Q}^{(\bw)}_k(u+w_k,\tau)=&{\sf
T}_{\bmu-u(e_k-e_{k-1})}\left(\frac{u}{N},\tau\right)=\notag\\
=&\tau^{-(N-1)H}
{\sf Q}_1(w_1,\tau)\mathcal{P}^{-1}\ldots
{\sf Q}_k(w_k+u,\tau)\mathcal{P}^{-1}\ldots\mathcal{P}^{-1}{\sf Q}_N(w_N,\tau)\,.
\end{align}
The definition is not unique in a sense that the Baxter operator depends on the arbitrary
parameters $\bw$ ($\bmu$). In the case of the
homogeneous spin chain there is a distinguished choice, $\bw=\bsigma$, which ensures that
$\mathcal{Q}_k(\sigma_k)=\mathcal{P}\tau^H$. However, it is clear that other variants
are also feasible~\footnote{
The choice $w\neq \sigma$ could  be useful for the analysis of the
homogeneous spin chains with a finite dimensional quantum space.
}.

New operators, $\mathcal{Q}^{(\bw)}_k(u,\tau)$, acting on the quantum space of the model,
form a commutative operator family. The normalization condition~(\ref{QPH}) which holds
in a homogeneous
case is replaced now by
\begin{align}\label{}
\mathcal{Q}^{(\bw)}_k(w_k,\tau)={\sf T}_{w}(0,\tau)\equiv \mathcal{Z}_w\,.
\end{align}
Using the commutation relations~(\ref{QcQ}) it is straightforward to derive
that the operators $\mathcal{Q}^{(\bw)}_k(u,\tau)$
possess all
the properties which hold in a homogeneous case. Namely, the generic transfer matrix
factorizes into the product of Baxter operators
\begin{align}\label{}
{\sf T}_{\brho}(u,\tau)=\mathcal{Z}_w^{1-N}\,\mathcal{Q}^{(\bw)}_1(u+\rho_1,\tau)\ldots
\mathcal{Q}^{(\bw)}_N(u+\rho_N,\tau)\,.
\end{align}
The fusion relation for the generic transfer matrices has exactly the same form  as in homogeneous case,
see Eq.~(\ref{TT-1})

\section{Summary}\label{summary}

In this paper we have developed an approach which allows to construct the Baxter
$\mathcal{Q}-$operators for a generic $sl(N)$ spin chain.
We have proven that the $sl(N)$ invariant
$\mathcal{R}-$operator on a tensor product of Verma modules
can be represented in  factorized form. The factorizing operators have an extremely
simple form~(\ref{r-kernel}) and possess a number of remarkable
properties~(\ref{propR}).
For the homogeneous spin chains we have defined the Baxter $\mathcal{Q}-$operators
as the trace of the monodromy matrix constructed of the factorizing operators.
We have shown that the Baxter $\mathcal{Q}-$operators
can be identified with  the transfer matrices for the
special choice of the auxiliary space.
This definition of the Baxter $\mathcal{Q}-$operators (see
Eq.~(\ref{Qw})  holds  for  inhomogeneous spin chains with impurities as well.

Many of  the properties of the Baxter $\mathcal{Q}-$operators and transfer matrices follow
readily from the properties of the factorizing operators, Eqs.~(\ref{propR}),~(\ref{Rz}).
In particularly, we have shown that the generic transfer matrix is factorized into the
product of $N$ different Baxter $\mathcal{Q}-$operators. This representation for the
transfer matrix together with the commutativity  of the Baxter $\mathcal{Q}-$operators results
immediately in certain functional relations for transfer matrices.

Another type of  fusion relations involves the transfer matrices with
a finite dimensional auxiliary space.
We recall, that   Verma module $\mathbb{V}_{\brho}$ has an
invariant finite dimensional submodule $\upsilon_{\brho}$ if
$\lambda_k=\rho_{k+1}-\rho_k+1=-n_k\leq 0$,  
$k=1,\ldots,N-1$. (~This representation  corresponds to the Young tableau 
specified by the partition $\{\ell_1,\ell_2,\ldots,\ell_{N-1}\}$, where 
$\ell_k=\sum_{i=k}^{N-1} n_i$ is the length of the $k-$th row in the tableau.~)
Let  $t_{\brho}(u,\tau)$ be a trace of the monodromy matrix over such finite dimensional spce
\begin{align}\label{tsmall}
t_{\brho}(u,\tau)=\tr_{\upsilon_\brho} \mathcal{R}_{10}(u,\tau)\ldots \mathcal{R}_{L0}(u,\tau)\,.
\end{align}
Assuming that the normalization of the factorizing operators are chosen
according to~Eq.~(\ref{AslN}) one can obtain the following determinant representation for 
 the transfer matrix~(\ref{tsmall}) 
\begin{align}\label{Qdet}
t_{\brho}(u,\tau)=\left(\mathcal{P}\tau^{H}\right)^{-N+1}\,
\left|\begin{array}{cccc} \mathcal{Q}_1(u+\rho_1,\tau)& \mathcal{Q}_1(u+\rho_2,\tau)&
\ldots&\mathcal{Q}_1(u+\rho_N,\tau)\\
\mathcal{Q}_2(u+\rho_1,\tau)& \mathcal{Q}_2(u+\rho_2,\tau)&
\ldots&\mathcal{Q}_2(u+\rho_N,\tau)\\
\vdots&\vdots&\ddots&\vdots\\
\mathcal{Q}_N(u+\rho_1,\tau)& \mathcal{Q}_N(u+\rho_2,\tau)&
\ldots&\mathcal{Q}_N(u+\rho_N,\tau)
\end{array}
\right|\,.
\end{align}
The proof of~(\ref{Qdet}) is based on the Berstein-Gel'fand-Gel'fand resolution 
of the finite-dimensional modules and will be given elsewhere~\cite{DMp}.
Equation~(\ref{Qdet}) gives rise to a variety of  functional
relations involving the Baxter $\mathcal{Q}-$operators and
(in)finite-dimensional transfer matrices.  The simplest of them
are the  so-called  Wronskian relation and Baxter equation.
One  easily derives from~Eq.~(\ref{Qdet}) the Wronskian relation, which in $\blambda$ notation, 
$t_{\brho}(u,\tau)\to
t_{\blambda}(u,\tau)=t_{\lambda_1\ldots\lambda_{N-1}}(u,\tau)$ reads
\begin{align}\label{Wr}
\left(\mathcal{P}\tau^{H}\right)^{N-1}\,t_{0\ldots 0}(u,\tau)\,=
\det|\mathcal{Q}_k(u+N-j,\tau)|_{k,j=1,\ldots,N}\,,
\end{align}
where the transfer matrix $t_{0\ldots 0}(u,\tau)$ is proportional to the unit operator on
quantum space.

Let us put
\begin{align}
t_0(u,\tau)=t_{N}(u,\tau)=t_{0\ldots 0}(u,\tau),\quad \text{and}\quad t_k(u,\tau)=
t_{0\ldots -1_k\ldots 0}(u,\tau)\,.
\end{align}
That is  the transfer matrix  $t_k(u,\tau)$  is given by  a trace of a monodromy matrix
over a finite dimensional auxiliary space  
which corresponds to the Young tableu with one column and $k$-rows.
 Following the lines of Ref.~\cite{DM-II} one can derive the
self-consistency equation (Baxter equation) involving the Baxter $\mathcal{Q}-$operators
and the finite
dimensional transfer matrices, $t_k(u,\tau)$.
It takes the form of the $N-$th order difference equation 
\begin{align}\label{BE}
\sum_{k=0}^{N}(-1)^k t_k(u+k/N,\tau)\, \mathcal{Q}_j(u+N-k,\tau)=0\,,
\end{align}
which, due to Eq.~(\ref{Q=T}),  can be considered as
a  fusion relation involving the  finite and infinite
dimensional transfer matrices of a special type.  Equation~(\ref{BE}) is a generalization
of the standard $sl(2)$ $T$-$Q$ relation.

Thus the 
operators $\mathcal{Q}_k(u)$ possess the following properties
\begin{itemize}
\item
Form a commutative family $[\mathcal{Q}_k(u),Q_j(v)]=0$
\item
Commute with all transfer matrices $[\mathcal{Q}_k(u),{\sf T}_\brho(v)]=0$
\item
Satisfy the $N-$th order difference equation~(\ref{BE}) infolving the 
finite dimensional transfer matrices.
\end{itemize}
The operators with such
properties are usually referred to as the Baxter $\mathcal{Q}-$operators, which justifies
using this name in the previous sections.

Let us notice that relations similar
to (\ref{Qdet}), (\ref{Wr}), (\ref{BE}) hold for
the spin chains with the affine $U_q(\widehat{sl}(n))$ symmetry
algebra~\cite{BHK,Kojima,K}~ (see also \cite{KLWZ,Z96}).
\vskip 5mm
It follows from Eqs.~(\ref{TQ}), (\ref{Qdet}) that the Baxter operators encode the full
information about the system. Provided that the eigenvalues of the Baxter operators are known
one can restore eigenvalues of all transfer matrices.
For low-rank symmetry models the knowledge of the eigenvalues of the Baxter operators
is sufficient to restore the wave function of the system. This can be done with the help
of the Separation of Variables method developed by E.~Sklyanin~\cite{Skl}. (Applications
for the specific models can be found in Refs.~\cite{SL2C,DKM-I,DKM-II,KL,KLS,KNS,S}.)

The  Hamiltonian of the system is defined as a logarithmic derivative of the
transfer matrix with the same representation in the quantum and auxiliary space.
It can be represented as the sum of pair-wise Hamiltonians
\begin{align}\label{Hm}
{\sf H}=\dfrac{d}{du}\log {\sf T}_{\bsigma}(u,\tau)\Big|_{u=0}
=\sum_{k=1}^L\tau^{H_{k}}\,\mathcal{H}_{kk+1}\,\tau^{-H_{k}}\,,
\end{align}
where
\begin{align}   \label{pwH}
\mathcal{H}_{kk+1}=\mathcal{R}_{kk+1}^{-1}(0) \dfrac{d}{du}\mathcal{R}_{kk+1}(u)|_{u=0}\,.
\end{align}
Using Eqs.~(\ref{RVV-factform}) and (\ref{r-kernel}) one easily finds the following
 expression for
the kernel of the pairwise Hamiltonian~(\ref{pwH})
\begin{align}\label{log}
\mathcal{H}(z,w|\alpha\,\beta)=
\log \left(\prod_{m=1}^{N} (\bar\beta_w w^{-1} z\bar\alpha^{-1}_z)_{mm}\right)\,
\mathcal{I}^{\bsigma}(z,\alpha)\mathcal{I}^{\bsigma}(w,\beta)\,.
\end{align}
Let us notice that for the $sl(2)$
case the argument of the logarithmic function turns into the invariant ratio
${(1+z\bar \beta)(1+w\bar\alpha)}/({(1+z\bar \alpha)(1+w\bar\beta)})$. (Here $z=z_{21}$,
$w=w_{21}$ etc.). The eigenvalues of the Hamiltonian~(\ref{Hm}) can be expressed in terms
of the eigenvalues of the Baxter operators as follows
\begin{align}\label{}
{\sf E}=\sum_{k=1}^L \frac{d}{du} \log \mathcal{Q}_k(u+\sigma_k)\Big|_{u=0}\,.
\end{align}
For a generic representation of the $sl(N)$ algebra it is not possible to define an invariant
scalar product  and the Hamiltonian~(\ref{Hm}) is not Hermitian in general. Of special
interest is a situation when the Verma modules possesses an invariant submodule
 which
admits an invariant scalar product. It could be a finite dimensional subrepresentation
which admits the $SU(N)$ invariant scalar product.
There are also  infinite dimensional invariant subspaces which can be equipped with
invariant scalar products. They can be identified with
unitary representations of the noncompact group $SU(m,N-m)$.
The spin chain with the $SU(2,2)$
symmetry group, for instance, is relevant for  a description of the anomalous dimensions
of a certain class of composite operators in Quantum Chromodynamics~\cite{BFHZ}.
We hope that the approach developed here will be useful for an analysis of this type of models.

\section*{Acknowledgment}
We are grateful to P.~P.~Kulish, M.~A.~Semenov-Tian-Shansky and
V.~O.~Tarasov for helpful discussions. This work was supported by
the RFFI grant 07-02-92166 (S.D., A.M.), RFFI grant 08-01-00638,
 DFG grant 436 Rus 17/4/07 (S.D.) and  by the German Research
 Foundation (DFG) grants 9209282 (A.M.).

\appendix
\renewcommand{\theequation}{\Alph{section}.\arabic{equation}}
\setcounter{table}{0}
\renewcommand{\thetable}{\Alph{table}}

\section{Appendix: The $SL(N,\mathbb{C})$ factorizing operator in the coherent state basis}
\label{lemma1}


We start the proof of    Eq.~(\ref{R-act}) with a remark that the coherent state~(\ref{cDD}) can be
represented in the form
\begin{align}\label{}
\Delta^{\bsigma\brho}(z,w|\alpha,\beta)=T^{\balpha}(\alpha^{-1})\,
T^{\bbeta}(\beta^{-1})\, \cdot 1\,.
\end{align}
Let us  apply the operator $\mathbb{R}^{(m)}$  to the function
\begin{align}\label{}
\Phi(z,w)=\int D\alpha D\beta \,f(\alpha,\beta) \,\Delta^{\bsigma\brho}(z,w|\alpha,\beta)\,,
\end{align}
where $f(\alpha,\beta)$ is a smooth function with a finite support.

The evaluation of $\mathbb{R}^{(m)}\Phi$ is based on the identity
\begin{align}\label{id-1}
\mathbb{R}^{(m)}(\lambda) T^{\balpha}(\alpha^{-1})\,
T^{\bbeta}(\beta^{-1})&
=\mathbb{W}_2\,
[(w^{-1}z)_{N1}]^\lambda\,\mathbb{W}_1 \,
T^{\balpha}(\alpha^{-1})\,
T^{\bbeta}(\beta^{-1})\notag\\
&=
T^{\balpha_{m,\lambda}}(\alpha^{-1})\,
T^{\bbeta_{m,-\lambda}}(\beta^{-1})\, \mathbb{W}_2\,
[(w^{-1}\beta\alpha^{-1}z)_{N1}]^\lambda\,\mathbb{W}_1\,,
\end{align}
where
\begin{align}
\mathbb{W}_1=&\left(
\overleftarrow{\prod_{j=m}^{N-1}} \mathbb{V}_j(\rho_{m,j+1})\right)\,
\left(
\overrightarrow{\prod_{i=1}^{m-1}} \mathbb{U}_i(\sigma_{im})\right)\,,\\
\mathbb{W}_2=&
\left(\overleftarrow{\prod_{i=1}^{m-1}} \mathbb{U}_i(\lambda-\sigma_{im})\right)
\left(\overrightarrow{\prod_{j=m}^{N-1}} \mathbb{V}_j(\lambda-\rho_{m,j+1})\right)\,.
\end{align}
To derive (\ref{id-1}) one uses the following identities,
\begin{align}\label{W1T}
\mathbb{W}_1\,T^{\balpha}(\alpha^{-1})\,
T^{\bbeta}(\beta^{-1})&=T^{\balpha'}(\alpha^{-1})\,
T^{\bbeta'}(\beta^{-1})\,\mathbb{W}_1\,,\\[2mm]
\label{zT}
[(w^{-1}z)_{N1}]^\lambda\, T^{\balpha'}(\alpha^{-1})\,
T^{\bbeta'}(\beta^{-1})=&
 T^{\balpha''}(\alpha^{-1})\,
T^{\bbeta''}(\beta^{-1})\,
[(w^{-1}\beta\alpha^{-1}z)_{N1}]^\lambda\,,\\
\label{W2T}
\mathbb{W}_2\,T^{\balpha''}(\alpha^{-1})\,
T^{\bbeta''}(\beta^{-1})&=T^{\balpha_{m,\lambda}}(\alpha^{-1})\,
T^{\bbeta_{m,-\lambda}}(\beta^{-1})\,\mathbb{W}_1\,,
\end{align}
where $\balpha''(h)=h_{11}^{-\lambda}\,\balpha'(h)$,~
$\bbeta''(h)=h_{NN}^{\lambda}\,\bbeta'(h)$ and
\begin{align}\label{}
\balpha'(h)=h_{11}^{\sigma_{1m}}\prod_{k=2}^m h_{kk}^{-\sigma_{k-1,k}}\,\balpha(h)\,,&&
\bbeta'(h)=h_{NN}^{-\rho_{mN}}\prod_{k=n}^{N-1} h_{kk}^{\rho_{k,k+1}}\,\bbeta(h)\,.
\end{align}
The identities~(\ref{W1T}) and (\ref{W2T}) follow directly from the intertwining relation~(\ref{UTT}).
To derive (\ref{zT}) it is sufficient to notice
\begin{align}\label{}
(w^{-1}_\beta \beta\alpha^{-1} z_\alpha)_{N1}=
(d_{w,\beta}\,\beta_{w}\, w^{-1}\, z\, \alpha_z^{-1} \,d^{-1}_{z,\alpha})_{N1}=
(d_{w,\beta})_{NN} (w^{-1} z)_{N1} (d^{-1}_{z,\alpha})_{11}\,,
\end{align}
where we recall that $\alpha\cdot z=z_\alpha\, d_{z,\alpha}\, \alpha_z$.

Let us now calculate   $\mathbb{W}_2\,
[(w^{-1}\beta\alpha^{-1}z)_{N1}]^\lambda\,\mathbb{W}_1$. To this end one can use the integral
representation for the intertwining operators~(\ref{prop}).
Let us show that
\begin{multline}\label{VSV}
\left(\overrightarrow{\prod_{j=m}^{N-1}} \mathbb{V}_j(\lambda-\rho_{m,j+1})\right)\,
[(w^{-1}\beta\alpha^{-1}z)_{N1}]^\lambda\,
\left(
\overleftarrow{\prod_{j=m}^{N-1}} \mathbb{V}_j(\rho_{m,j+1})\right)
=r_m(\lambda)\, [(w^{-1}\beta\alpha^{-1} z)_{m1}]^{\lambda}\,\,,
\end{multline}
where
\begin{align}\label{}
r_m(\lambda)=\prod_{j=m+1}^N\frac{A(\lambda-\rho_{mj})}{A(-\rho_{mj})}=
\prod_{j=m+1}^N\frac{A(u_m-v_j)}{A(v_m-v_j)}\,,
\end{align}
and the parameters $u_k,v_k,\lambda, \sigma_k,\rho_k$ are related by~(\ref{rel-uv}).
We represent l.h.s. of (\ref{VSV}) in the form
\begin{multline}\label{Int-f}
\prod_{j=m+1}^{N}\,A(\rho_{mj})A(\rho_{jm}+\lambda)
\int d^{2}\xi'_{j-1}\,\int
d^{2}\xi_{j-1}\\
[\xi_{j-1}]^{-(1+\lambda+\rho_{jm})}\,[\xi'_{j-1}-\xi_{j-1}]^{-(1+\rho_{mj})}
\,[ \left(w_\xi^m)^{-1}\beta\alpha^{-1} z\right)_{N1}]^{\lambda}\,,
\end{multline}
where
\begin{align}
w_\xi^m=&w\left(1-\xi_{m}e_{m+1m}\right)
\ldots\left(1-\xi_{N-1}e_{NN-1}\right)\,.
\end{align}
It follows from~(\ref{id-1}) that the integrations in Eq.~(\ref{Int-f})
have to be carried out in the following order
$\xi_{m},\ldots,\xi_{N-1}, \xi'_{N-1},\ldots,\xi'_{m}$. Since the integrals
converge, we rearrange them and carry out
the integration over $(\xi_{N-1},\xi'_{N-1})$, then over $(\xi_{N-2},\xi'_{N-2})$,
and so on till $(\xi_m,\xi'_m)$.
Each integration is reduced to the standard integral %
\begin{align}\label{}
\int d^2\xi' \int d^2\xi\,{[\xi]^{-1-\lambda+\rho}\,[\xi'-\xi]^{-1-\rho}\,
[\xi-x_0]^{\lambda}}=\frac1{A(\rho)A(-\rho)}\,.
\end{align}
(We recall here that $[a]^\lambda=a^\lambda (a^*)^{\bar\lambda}$.)
Then collecting all factors and taking into account that $(e_{NN-1}\ldots e_{m+1m}
w^{-1}\beta\alpha^{-1}z)_{N1}=(w^{-1}\beta\alpha^{-1}z)_{m1}$ one gets Eq.~(\ref{VSV})\,.
\vskip 5mm

Quite similarly one obtains
\begin{align}\label{}
\left(\overleftarrow{\prod_{i=1}^{m-1}} \mathbb{U}_i(\lambda-\sigma_{im})\right)
\,[(w^{-1}\beta\alpha^{-1}z)_{m1}]^\lambda\,
\left(
\overrightarrow{\prod_{i=1}^{m-1}} \mathbb{U}_i(\sigma_{im})\right)
=p_m(\lambda)[(w^{-1}\beta\alpha^{-1}z)_{mm}]^\lambda\,,
\end{align}
where  %
\begin{align}\label{}
p_m(\lambda)=\prod_{k=1}^{m-1}(-1)^{\lambda-\bar\lambda}\frac{A(\lambda-\sigma_{km})}{A(-\sigma_{km})}
=\prod_{k=1}^{m-1}(-1)^{\lambda-\bar\lambda}\frac{A(u_k-v_m)}{A(u_k-u_m)}\,.
\end{align}

Using these results one obtains from~(\ref{id-1})
\begin{align}\label{}
\mathbb{R}^{(m)}(\lambda) \Delta^{\bsigma\brho}(z,w|\alpha,\beta)=&
f_{\bsigma\brho}^{(m)}(\lambda)\,T^{\balpha_{m,\lambda}}(\alpha^{-1})\,
T^{\bbeta_{m,-\lambda}}(\beta^{-1})\, [(w^{-1}\,\beta\,\alpha^{-1}\,z)_{mm}]^\lambda\notag\\
=&f_{\bsigma\brho}^{(m)}(\lambda)\,
[(\beta_w\, w^{-1}\,z\,\alpha_z^{-1})_{mm}]^{\lambda}\,\Delta^{\bsigma\brho}(z,w|\alpha,\beta)\,.
\end{align}
%

\section{Appendix: Proof of the relations~(\ref{DR}) and (\ref{p1}) }\label{AppB}
%
In this Appendix we give the proof of the relations~(\ref{DR}) and (\ref{p1}) for the
factorizing operators~$\mathbb{R}^{(m)}$.
We start from Eqs.~(\ref{DR}) and   represent
the operator $D_{k+1,k}$ in the form (see Eq.~(\ref{EzD}))
\begin{align}\label{}
D_{k+1,k}=-\sum_{in}\,(z^{-1})_{ki}\, z_{n,k+1}\, E_{ni}\,.
\end{align}
Since the relations~(\ref{DR}) hold for the $SL(N,\mathbb{C})$ factorizing operators
one can derive the following equation for the kernel
$\mathcal{K}_{\lambda,\bsigma,\brho}^{(m)}(z,w|\alpha,\beta)$,
\begin{align}\label{DDR}
D^{(z)}_{k+1,k}
\mathcal{K}_{\lambda,\bsigma,\brho}^{(m)}(z,w|\alpha,\beta)=
\sum_{in}\,(z^{-1})_{ki}\, z_{n,k+1}
\widetilde E^{(\alpha)}_{ni}\,\mathcal{K}_{\lambda,\bsigma,\brho}^{(m)}(z,w|\alpha,\beta)\,,
\end{align}
where $k>m+1$ and we have used the commutation relations~(\ref{Rz}) for the $SL(N,\mathbb{C})$
operators. Taking into account that the kernel of the $sl(N)$ factorizing
operator $\mathbb{R}^{(m)}(\lambda)$ is given by the function
$\mathcal{K}_{\lambda,\bsigma,\brho}^{(m)}(z,w|\bar\alpha,\bar\beta)$
and making use of Eqs.~(\ref{EtE}) one can easily check that
Eq.~(\ref{DDR})  gives rise to the following equation
\begin{align}\label{}
D^{(z)}_{k+1,k}\,\mathbb{R}^{(m)}(\lambda)=\mathbb{R}^{(m)}(\lambda)\,D^{(z)}_{k+1,k}
\end{align}
for the $sl(N)$ factorizing operator $\mathbb{R}^{(m)}(\lambda)$.

\vskip 3mm

The  relation~(\ref{p1}) is equivalent to the following equation for the kernels
\begin{align}\label{RK}
\mathcal{R}^{(m)}_{\lambda+\mu,\bsigma\brho}(\eta,\xi|\bar\alpha,\bar\beta)=
\Omega_{\bsigma'\brho'}\left(\mathcal{R}^{(m)}_{\mu,\bsigma'\brho'}(\eta,\xi|\bar z,\bar w)
\mathcal{R}^{(m)}_{\lambda,\bsigma\brho}(z,w|\bar \alpha,\bar \beta)\right)\,.
\end{align}
This equation  is a consequence    of   the commutation relations~(\ref{Rz}).

Let us rewrite the kernel $\mathcal{R}^{(m)}_{\lambda,\bsigma\brho}(z,w|\bar \alpha,\bar
\beta)$
as follows
(for  brevity we will omit the normalization factor $A_m$)
\begin{align}\label{tran}
\left(\bar\beta_w\,w^{-1}\, z\, \bar\alpha_z^{-1}
\right)_{mm}^\lambda\,
\mathcal{I}^\bsigma(z,\alpha)\,\mathcal{I}^\brho(w,\beta)=
\left(w_\beta^{-1}\,\bar\beta^{-1}\, \bar\alpha^{-1}z_\alpha
\right)_{mm}^\lambda\,
\mathcal{I}^{\bsigma'}(z,\alpha)\,\mathcal{I}^{\brho'}(w,\beta)\,.
\end{align}
The key point is that the  factor
$X_m=\left(w_\beta^{-1}\,\bar\beta^{-1}\, \bar\alpha^{-1}z_\alpha\right)_{mm}$ depends
only on
the variables $z_{jk}^{-1}$, with $j> m$ and on the variables $w_{kj}$ with $j<m$.
It can be shown as follows (we consider only $z$ case): $X_m$ depends on the variables
$(z_\alpha)_{km}$ with $k=m+1,\ldots,N$, which, due to the triangularity of the matrix
$z_\alpha$, are given by linear combinations of the elements $(z_\alpha^{-1})_{kp}$ with
$p\geq m$. The matrix $z_\alpha^{-1}$ is determined by the equation
\begin{align}\label{a-z-m}
z^{-1}\,\bar\alpha^{-1}=\bar\alpha_z^{-1}\,d_{z,\alpha}^{-1}\, z_\alpha^{-1}\,,
\end{align}
which follows from Eq.~(\ref{alphaz}). It is straightforward to see from the above
equation that the matrix elements $(z_\alpha^{-1})_{kn}$ depend on the elements
$z_{jn}^{-1}$ with $j\geq k$ only.
Therefore, taking into account the commutation relations  of Eq.~(\ref{Rz}) one can
transform~(\ref{RK}) to
\begin{multline}\label{om1}
\Omega_{\bsigma'\brho'}\left(\mathcal{R}^{(m)}_{\mu,\bsigma'\brho'}(\eta,\xi|\bar z,\bar w)
\left(\xi_\beta^{-1}\,\bar\beta^{-1}\, \bar\alpha^{-1}\eta_\alpha
\right)_{mm}^\lambda\,
\mathcal{I}^{\bsigma'}(z,\alpha)\,\mathcal{I}^{\brho'}(w,\beta)\right)=\\
=\left(\xi_\beta^{-1}\,\bar\beta^{-1}\, \bar\alpha^{-1}\eta_\alpha
\right)_{mm}^\lambda\Omega_{\bsigma'\brho'}\left(\mathcal{R}^{(m)}_{\mu,\bsigma'\brho'}(\eta,\xi|\bar
z,\bar w)
\mathcal{I}^{\bsigma'}(z,\alpha)\,\mathcal{I}^{\brho'}(w,\beta)\right)\\=
\left(\xi_\beta^{-1}\,\bar\beta^{-1}\, \bar\alpha^{-1}\eta_\alpha
\right)_{mm}^\lambda\mathcal{R}^{(m)}_{\mu,\bsigma'\brho'}(\eta,\xi|\bar
\alpha,\bar \beta)=\mathcal{R}^{(m)}_{\mu+\lambda,\bsigma\brho}(\eta,\xi|\bar
\alpha,\bar \beta)\,,
\end{multline}
where on  the last step we use the transformation~(\ref{tran}).

\section{Appendix: Proof of the estimate~(\ref{estR}) for matrix 
                   elements of  factorizing operators }
\label{R-asymptotic}
Here we  derive the estimate~(\ref{estR}) for the matrix element of the factorizing operator
$\mathcal{R}^{(k)}_{10}(\lambda)$. Let $e_n(z)$ and $e_k(w)$ be  the basis
vectors~(\ref{enz}) in the quantum and auxiliary spaces, respectively.
Defining
\begin{align}\label{}
E_n(z)=\prod_{i>k}\frac{z^{n_{ik}}}{n_{ik}!}
\end{align}
one can represent $e_n(z)$ as
\begin{align}\label{}
e_n(z)=\Omega_{kn}(\bsigma)
 E_k(\partial_{\bar\alpha}) \mathcal{I}^\bsigma(z,\alpha)|_{\alpha=1 (\alpha_{ij}=0, i>j)}\,,
\end{align}
where $\Omega_{kn}(\bsigma)\equiv\Omega_\bsigma(\bar e_k,e_n)$ and the sum over repeating
(multi)indices is
implied.
The matrix element $[\mathcal{R}^{(k)}_{10}(\lambda)]_{mn}^{m'n'}$  can be expressed as
follows
\begin{align}\label{EEE}
[\mathcal{R}^{(k)}_{10}(\lambda)]_{mn}^{m'n'}=
\Omega_{ni}(\brho)\,\Omega_{mj}(\bsigma)\,
E_{m'}(\partial_{z})\, E_{n'}(\partial_w)\, E_{j}(\partial_{\bar\alpha})\,E_{i}(\partial_{\bar\beta})
\mathcal{R}^{(m)}_{\lambda,\bsigma\brho}(w,z|\bar\alpha,\bar\beta)\Big|_{z=w=\alpha=\beta=1}\,.
\end{align}
We are interested in the behavior of this matrix element at $n_{jk}\to\infty$,
$j=k+1,\ldots,N$, all other indices $m_{ji}, m'_{ji}$ and $n_{ji}$, $n'_{ji}$, $i\neq
k$ being fixed. It is clear that the multi-index $j$ in r.h.s. of (\ref{EEE}) varies in
finite limits. Using Eq.~(\ref{r-kernel}) for $\mathcal{R}^{(m)}_{\lambda,\bsigma\brho}$
and carrying out differentiation with respect to $\bar\alpha$ and $z$ one finds that the
matrix element~(\ref{EEE}) is given by a sum of terms which have the form
\begin{align}\label{E2}
\Omega_{ni}(\brho)\,
E_{n'}(\partial_w)\,E_{i}(\partial_{\bar\beta})
(\bar\beta w)_{kk}^{\lambda-K}\, P(w,\bar\beta)\Big|_{w=\beta=1}\,,
\end{align}
where $K$ is some constant.
The polynomial $P(w,\bar\beta)$ has a finite degree and does not contain large factors
$n_{jk}$.
The factor $(\bar\beta w)_{kk}=1+\sum_{p} \bar\beta_{pk} w_{pk}$ depends only on the
``large'' variables $w_{pk}$,~$\bar\beta_{pk}$. After differentiation with respect to all other
variables one gets for~(\ref{E2})
\begin{align}\label{E3}
\Omega_{ni}(\brho)\,
\prod_{p=k}^N \frac1{n'_{pk}!}\frac1{i_{pk}!}
\left(\frac{\partial\phantom{w}}{\partial{w_{pk}}}\right)^{n'_{pk}}
\left(\frac{\partial\phantom{w}}{\partial{\bar\beta_{pk}}}\right)^{i_{pk}}
(\bar\beta w)_{kk}^{\lambda-K}\, \widetilde P(w_{pk},\bar\beta_{pk})\Big|_{w_{pk}=\beta_{pk}=0}\,,
\end{align}
where again the polynomial $\widetilde P$ does not contains large factors $n_{jk}$.
Finally, one gets that~(\ref{E3}) is given by the (finite) sum of the terms
\begin{align}\label{E4}
\Omega_{ni}(\brho)\,\frac{\Gamma(-\lambda+K+\sum_{j=k+1}^N(n'_{jk}-s_j))}{(n'_{k+1k}-s_{k+1})!\ldots
(n'_{Nk}-s_N)!}
\times R(n'_{k+1k},\ldots,n'_{Nk})\,,
\end{align}
where $s_j$ are some constants, $R$ is a polynomial and the difference of indices
$i_{jk}-n'_{jk}$ is finite at $n'_{jk}\to \infty$, $j=k+1,\ldots,N$.
\vskip 5mm

Let us estimate the coefficient $\Omega_{ni}(\brho)$ when all  indices
except $n_{jk}$ and $i_{jk}$, $j>k$ are small. To this end we will use the
integral representation~(\ref{def-mu}),
\begin{align}\label{OmI}
\Omega_{ni}(\brho)=c_N(\brho)\int Dz\, \mu_{N}(z)\, \overline{e_n(z)}\, e_i(z)\,,
\end{align}
with $\mu_N(z)=\prod_{j=1}^{N-1}\Delta_j^{-\lambda_j-1}(z^\dagger z)$, $\lambda_j=\rho_j-\rho_{j+1}$.
We recall that by definition the integral~(\ref{OmI})  for arbitrary $\{\lambda_j\}$ is
understood as an analytic continuation from the region of $\lambda'$s where it converges.
To stress that the matrix element $\Omega_{ni}(\brho)$ is considered in the special
``kinematic'' ($n_{jk},j_{jk}\to\infty$)
 we put  more labels on it, $\Omega_{ni}(\brho)\to \Omega^{N,k}_{ni}(\blambda)$,
and switch from $\brho$ to $\blambda$.

To get the necessary estimate we proceed as follows:
 First we show  that the coefficient
$\Omega^{N,k}_{ni}(\blambda)$   for $k>1$ can be represented as the sum of  integrals~(\ref{OmI})
for $sl(N-1)$ case
\begin{align}\label{ind-1}
\Omega^{N,k}_{ni}(\blambda)=\sum_{q} c_q\, \Omega^{N-1,k-1}_{n_q,i_q}(\blambda_q)\,.
\end{align}
It is important that   the sum in (\ref{ind-1}) over $q$ goes in a finite range and
the  coefficients  $c_q$ and  parameters $(\lambda_q)_j=\lambda_j+\delta_j$
do not depend on $n_{jk},i_{jk}$. Continuing this procedure
one  can represent $\Omega^{N,k}_{ni}(\blambda)$ as the sum of the elements
$\Omega^{M,1}_{n_q i_q}(\blambda_q)$, where $M=N-k+1$, thus reducing the problem to
an evaluation of the matrix elements $\Omega^{M,1}_{n_q i_q}(\blambda_q)$.

To prove~(\ref{ind-1}) let us make the change of variables $z=w^{-1}$ and carry out the
integrations over variables $w_{21},\ldots,w_{N1}$. Let us represent matrix $w$ as
\begin{align}\label{wp}
w=\begin{pmatrix}1 &\vec{0}\\
                 \vec{a}&b\end{pmatrix}=
\begin{pmatrix}1&0&0&\ldots&0\\
a_1&1&0&\ldots&0\\
a_2&b_{21}&1&\ldots&0\\
\vdots&\vdots&\vdots&\ddots&\vdots\\
a_{N-1}&b_{N-11}&\ldots&b_{N-1,N-2}&1
\end{pmatrix}
\end{align}
and examine dependence of  $\Delta_p$ on the elements $a_1,\ldots,a_{N-1}$.
For $N\times N$ matrix $M$ we define
$\widetilde \Delta_p(M)=\det M_p$ and  $M_p$ is $(N-p)\times(N-p)$ matrix
with elements $M_{ij}$, $i,j=p+1,\ldots,N$.
Noticing that
  $\Delta_p(z^\dagger z)=\widetilde \Delta_p(w w^\dagger)=\det |(bb^\dagger)_{ij}+a_i\bar
a_j|_{i,j=p,\ldots N-1}$ one concludes that
\begin{align}
\Delta_1=\Delta_1(a_{1},\ldots,a_{N-1})\,,&&
\Delta_2=&\Delta_2(a_{2},\ldots,a_{N-1})\,,&&
\ldots&&
\Delta_{N-1}=\Delta_{N-1}(a_{N-1}).
\end{align}
Let $B=bb^\dagger$ and $A_p=(a_{p},\ldots, a_{N-1})$.
Taking into account that
\begin{align}\label{}
\widetilde \Delta_{p}(w w^\dagger)=\det(B_{p-1}+A_p\otimes A_p^\dagger)=
\widetilde \Delta_{p-1}(bb^\dagger)\cdot (1+A^\dagger_p\, B_{p-1}^{-1}\, A_p)\,
\end{align}
one gets for the measure
\begin{align}\label{}
\mu_N(w)=&\prod_{p=1}^{N-1}\widetilde \Delta_p^{-\lambda_p-1}(ww^\dagger)=
\prod_{p=1}^{N-1}\widetilde \Delta_{p-1}^{-\lambda_{p}-1}(bb^\dagger) \,
(1+A^\dagger_p\, B_{p-1}^{-1}\, A_p)^{-\lambda_{p}-1}
\notag\\
=& \mu_{N-1}(b)\,
\prod_{p=1}^{N-1}(1+A^\dagger_p\, B_{p-1}^{-1}\, A_p)^{-\lambda_{p}-1}\,,
\end{align}
where
$
\mu_{N-1}(b)=\prod_{p=1}^{N-2} \widetilde \Delta_p^{-\lambda_{p+1}-1}(bb^\dagger)\,
$
~($\widetilde\Delta_0(bb^\dagger)=1$).
Now one can consequently carry out the integration over $a_1,a_2,\ldots,a_{N-1}$.
Indeed, let us consider integral
\begin{align}\label{int-a}
\int d^2 a_1 \, a_1^m \, \bar a_1^{\bar m}\,\left(1+A_1^\dagger\, B^{-1}\,
A_1\right)^{-\lambda_1-1}\,.
\end{align}
Representing $B^{-1}_{ij}=m_{ij}/\det B$, where
$m_{ij}=(-1)^{i+j} M_{ij}$, $M_{ij}$ being a minor of the matrix $B$, one derives
\begin{align}\label{}
\left(1+A^\dagger_1\, B^{-1}\, A_1\right)=&
\left(1+\frac1{\det B}\left[
|a_1|^2 m_{11}+a_1^\dagger m_{1j} a_j+a_i^\dagger m_{i1}a_1+ a^\dagger_i m_{ij} a_j
\right]\right)\notag
\\
=&
\left(1+\frac{m_{11}}{\det B}\left[
|a_1+m_{1j}/m_{11} a_j|^{2}+
\frac{1}{m_{11}^2}a^\dagger_i \left\{m_{11} m_{ij}- {m_{i1}m_{1j}}\right\}a_j
\right]\right)\notag\\
=&
\left(1+\frac{m_{11}}{\det B}
|a_1+m_{1j}/m_{11} a_j|^{2}+a^\dagger_i (B_1^{-1})_{ij} a_j
\right)
\,.
\end{align}
Here the summation over repeated indices ( $i,j=2,\ldots,N-1$ ) is implied and we make use of
the identity
\begin{align}\label{}
\left\{m_{11} m_{ij}- {m_{i1}m_{1j}}\right\}=m_{11}\det B \cdot (B^{-1}_{1})_{ij}\,.
\end{align}
After shifting and rescaling the integration variable  one gets for~(\ref{int-a})
\begin{align}\label{}
\sum_{q=0}^{\min(m,\bar m)}c_q(\lambda_1)\left(\frac{\det B}{m_{11}}\right)^{q+1}
 \left(-\frac{m_{1j} a_j}{\det B}\right)^{m-q}\,
\left(-\frac{\bar m_{1j} \bar a_j}{\det B}\right)^{\bar m-q}
(1+A_{2}^\dagger B_{1}^{-1} A_2)^{-\lambda_1+q}\,,
\end{align}
where
$c_q(\lambda)= C^m_q C^{\bar m}_q {q!\Gamma(\lambda-q)}/{\Gamma(\lambda+1)}$
and $m_{11}=\det B_1=\widetilde \Delta_1(bb)^\dagger$. Evidently, integrating over $a_2,a_3,\ldots$
one again encounters the  integrals of the type~(\ref{int-a}), hence the final
result of integrations  can be cast into the form~(\ref{ind-1}). This
calculation also shows that the integral~(\ref{OmI}) is a meromorphic function of
$\lambda_1,\ldots,\lambda_{N-1}$.

\vskip 5mm

In order to  calculate the element $\Omega^{N,1}_{ni}(\lambda_q)$
one can  repeatedly integrate  over the variables $z_{Nj}$, $j>1$ in the last row,
then  over $z_{N-1j}$, $j>1$ in the row $N-1$ and
so on. Representing the matrix $z$ in the form
\begin{align}\label{}
z=\begin{pmatrix}1&0&0&\ldots&0\\
b_{21}&1&0&\ldots&0\\
\vdots&\vdots&\ddots&&\vdots\\
b_{N-11}&b_{N-12}&\ldots&1&0\\
a_{1}&a_{2}&\ldots&a_{N-1}&1
\end{pmatrix}\,,
\end{align}
$(z^\dagger z)_{ij}=(b^\dagger b)_{ij}+\bar a_i a_j$, for $i,j=1,\ldots,
N-1$, one easily finds that $\Delta_k(z)$ depends only on the variables $a_1,\ldots,a_k$.
The integration over $a_{N-1},a_{N-2}\ldots, a_2$  goes along the same
lines as before. For instance,
$$
\prod_{k=2}^{N-1}\int d^2a_k \prod_{k=2}^{N-1}\Delta_{k}^{-\lambda_k-1}(z^\dagger z)=
\gamma_{N-1}\left[ \prod_{k=2}^{N-2}\Delta_{k}^{-\lambda_k-1}(b^\dagger b)\right]
R(z_{21},\ldots,z_{N1})
$$
where
$
\gamma_{N-1}=\pi^{N-2}(\lambda_{N-1}(\lambda_{N-1}+\lambda_{N-2})\ldots
(\lambda_{N-1}+\ldots+\lambda_2))^{-1}
$
and
$$
R(z_{21},\ldots,z_{N1})=
\frac{\left(1+
|z_{21}|^2+\ldots+|z_{N-11}|^2\right)^{\lambda_2+\ldots+\lambda_{N-1}-1}}
{\left(1+
|z_{21}|^2+\ldots+|z_{N1}|^2\right)^{\lambda_2+\ldots+\lambda_{N-1}}}\,.
$$
Therefore, one gets
\begin{multline}\label{}
\left( \prod_{1<j<k\leq N-1} \int d^2z_{kj} \right)\mu_N(z) \,\overline{e_n(z)}\,{e_i(z)}=\\
\sum_q c(\lambda,q)\,\, \prod_{p=2}^N z_{p1}^{m_{p}}\,\bar z_{p1}^{l_{p}}\,
\,\prod_{n=2}^{N-1}\left(1+\sum_{k=2}^n|z_{k1}|^2\right)^{\mu_n-1}
\left(1+\sum_{k=2}^N|z_{k1}|^2\right)^{-\Lambda_N-1}\,,
\end{multline}
where $m_{p}=i_{p1}+r_p(q)$, $l_p=n_{p1}+s_{p}(q)$, $\mu_p=\lambda_p+\delta_p(q)$ and
$\Lambda_N=\mu_1+\ldots+\mu_{N-1}$. The parameters $c(\lambda,q),r_p(q),s_{p}(q),\delta_p(q)$
do not depends on $n_{k1}, i_{k1}$, $k=2,\ldots,N$ and the over multi-index $q$ goes in the
finite limits.
Finally, after integration over $z_{p1}$, $p=2,\ldots,N$
one gets for each term in the sum
\begin{align}\label{}
\pi^{N-1} c(\lambda,q)\prod_{p=2}^N (-1)^{m_p+1}
\delta_{m_p,l_p}
m_p!\prod_{k=3}^{N}\frac{\Gamma(1-\Lambda_{k-1}+M_{k})}{\Gamma(1-\Lambda_{k}+M_k)}
\,\frac{\Gamma(1-\Lambda_{N})}{\Gamma(1-\Lambda_{2}+M_{2})}\,,
\end{align}
where $\Lambda_k=\mu_1+\ldots+\mu_{k-1}$ and $M_k=m_k+\ldots+m_N$. Together with
Eq.~(\ref{E4}) this results in the estimate~(\ref{estR}).

\end{document}